\documentclass[a4paper,11pt]{article}
\pdfoutput=1 

\usepackage{jheppub} 

\usepackage[T1]{fontenc} 

\usepackage{xspace}

\usepackage[normalem]{ulem}
\usepackage{multirow}
\usepackage{bashful}
\usepackage{todonotes}
\usepackage{subcaption}
\usepackage{feynmp-auto}

\newcommand{\LiteRed}{\texttt{LiteRed}\xspace}
\newcommand{\Libra}{\texttt{Libra}\xspace}

\newcommand{\egeXX}{\ensuremath{e^-\gamma \to e^-X\bar{X}}\xspace}
\newcommand{\egeee}{\ensuremath{e^-\gamma \to e^-e^-e^+}\xspace}
\newcommand{\egegg}{\ensuremath{e^-\gamma \to e^-\gamma\gamma}\xspace}

\newcommand{\egemumu}{\ensuremath{e^-\gamma \to e^-\mu^+ \mu^-}\xspace}

\newcommand{\A}{\mathcal{A}}

\newcommand{\e}{\ensuremath{\epsilon}}
\renewcommand{\Im}{\mathop{\mathrm{Im}}\nolimits}

\newcommand{\bs}{\boldsymbol}

\title{Total cross sections of $e\gamma\to e X\bar{X}$ processes with $X=\mu,\gamma, e$ via multiloop methods.}

\author[a]{Roman N. Lee,}
\author[a,b]{Alexey A. Lyubyakin}
\author[a,b]{and Vyacheslav A. Stotsky}

\affiliation[a]{Budker Institute of Nuclear Physics, 630090, Novosibirsk, Russia}
\affiliation[b]{Novosibirsk State University, 630090, Novosibirsk, Russia}

\emailAdd{r.n.lee@inp.nsk.su}
\emailAdd{a.liubiakin@g.nsu.ru}
\emailAdd{stotsky.slava@gmail.com}

\abstract{Using modern multiloop calculation methods, we derive the analytical expressions for the total cross sections of the processes $e^-\gamma \to e^-X\bar{X}$ with $X=\mu,\,\gamma$ or $e$ at arbitrary energies. For the first two processes our results are expressed via classical polylogarithms. The cross section of $e^-\gamma \to e^-e^-e^+$ is represented as a one-fold integral of complete elliptic integral $\mathrm{K}$ and logarithms. Using our results, we calculate the threshold and high-energy asymptotics and compare them with available results.}

\begin{document} 
\maketitle
\flushbottom

\section{Introduction}

Since the invention of quantum electrodynamics (QED), one of its first touchstones was the calculation of the cross sections of elementary processes, like $e^+e^-\to \mu^+\mu^-$ or $e\gamma\to e\gamma$. In particular, their total Born cross sections for arbitrary energies invariably appear in any QED textbook. In its vast majority, these results concern the processes with two particles in the initial state and two in the final ($2\to2$ processes). In contrast, the total Born cross sections of the $2\to N$ processes with $N>2$ have been paid much less attention\footnote{Perhaps, the only remarkable exception is the Racah results \cite{Racah1934a,Racah1934} for the processes $\gamma Z\to e^+e^- Z$ and  $e^-Z\to e^-\gamma Z$.}. This circumstance is not incidental. It appears that, when there are massive particles in the final state and/or massive propagators in the amplitude, the $N$-particle phase-space integrals are not so simple to be taken by brute force. From the viewpoint of contemporary multiloop methods, this is no wonder, as the $N$-particle phase space integral corresponds to the $N-1$-loop momentum space integral with bipartite cut. Thus, $N\geqslant 3$ case corresponds to $(L\geqslant 2)$-loop integrals, which are known to bring much more complexity than one-loop integrals. However, the last 40 years of development of multiloop calculations methods have passed for good reason, and we are now in perfect position to patch this omission. We should not expect too simple results though. It is well-known that already two-loop integrals can be impossible to express via harmonic polylogarithms \cite{Remiddi:1999ew} and even via generalized polylogarithms \cite{Goncharov1998}.

In the present paper we calculate the total cross sections of the processes \egeXX with $X=e^-,\mu^-,\gamma$. These processes have many astrophysical applications, see, e.g., Refs. \cite{1991A&A...252..414D,ENDO1993517,10.1093/mnras/266.4.910,PhysRevLett.86.1430,PhysRevD.64.071302,haug2004pair,Ravenni:2020ven}, and have been considered in a number of papers  \cite{votruba1948pair,mork1967pair,PhysRevA.4.917,Baier,haug1981simple,gould1984cross,ENDO1993517,anguelov1999numerical,PhysRevD.64.071302}. 
Our approach is based on using the optical theorem and Cutcosky rules to express the total cross section via cut diagrams in forward kinematics. The master formula is
\begin{equation}
    \sigma_{\egeXX}=\frac{\Im \A_{\egeXX}}{s-m_e^2},\label{intro:ot}
\end{equation} 
where $s=(p_1+k_1)^2$ is the square of total c.m. energy, $m_e$ is the electron mass, and $\Im \A_{\egeXX}$ is the sum of the propagator-type diagrams with cut $e^-,\ X,\ \bar{X}$ lines. Thus, in order to calculate the cross section, we apply to this sum the contemporary multiloop calculation methods: the IBP reduction and the calculation of the master integrals via the differential equations method.

Our calculation features a few methods not so widely known in the multiloop community. First, given two differential systems, $\partial_x \bs j=M_x\bs j$ and  $\partial_y \bs j=M_y\bs j$, we consider the solution of the second system in the form of generalized power series $\bs j=\sum_{\alpha\in S+\mathbb{Z}} \bs c_\alpha(x) y^\alpha$ and show how to systematically obtain the differential equations with respect to $x$ for the finite subset $\{\bs c_\alpha(x)|\alpha\in S\}$. Next, we demonstrate that the contribution of soft photons, as defined by the soft-photon approximation, can be calculated by means of the multiloop techniques, although it does not directly correspond to a sum of conventional Feynman diagrams. Finally, we successfully apply the recently introduced approach to the non-polylogarithmic integrals, based on the construction of $\e$-regular basis, Ref. \cite{Lee2019}.

\section{Total cross section of the process $e^-\gamma \to e^-\mu^+\mu^-$}

\begin{figure}
    \centering
    \includegraphics[width=0.83\linewidth]{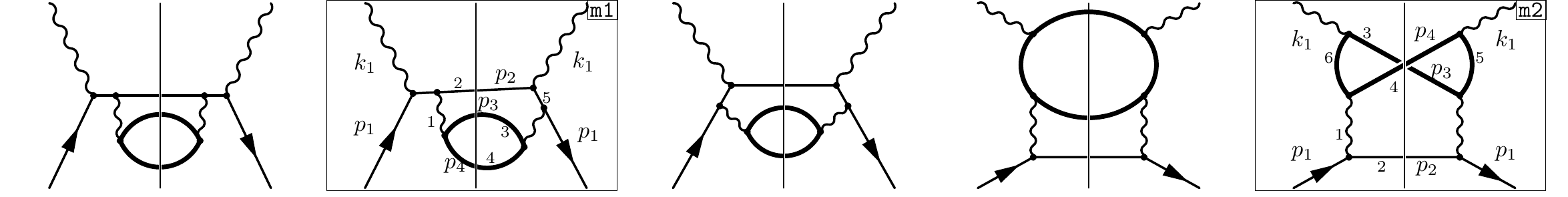}
    \caption{Diagrams contributing to the total cross section of the \egemumu process. First three diagrams correspond to $C$-odd muon pair, while the last two diagrams --- to $C$-even pair.}
    \label{fig:egemumudiagrams}
\end{figure}
Let us first consider the process $\egemumu$. As the muon mass $m_\mu$ is much larger than $m_e$, we can treat $m_e$ as a small parameter. Nevertheless, we can not simply put $m_e=0$ because the cross section becomes infrared divergent in this limit. Therefore, we keep the electron mass in ``large'' logarithms and omit the power corrections in $m_e$, so that our final result has the following form:
\begin{equation}
    \sigma_{\egemumu}=\sigma_0+\sigma_1 \ln (m_\mu/m_e).
\end{equation}
Below in this section we will use the units $m_{\mu}=1$.
In order to obtain the above form, we proceed in the following way. We define two \LiteRed \cite{Lee2013a} bases \texttt{m1} and \texttt{m2} containing the denominators of the corresponding boxed diagrams in Fig. \ref{fig:egemumudiagrams}. We add the irreducible numerators $D_6=k_{1}\cdot p_{3}$ and $D_7=p_{1}\cdot p_{3}$ to basis \texttt{m1} and $D_7=p_{1}\cdot p_{3}$  to \texttt{m2}.   
We reveal $7$ master integrals
\begin{gather*}
    \boldsymbol{\mathrm{j}}^\intercal=(j_1,j_2,...,j_7)=\left(j_{0111000}^{\texttt{m1}},\,j_{-1111000}^{\texttt{m1}},\,j_{-2111000}^{\texttt{m1}},\,j_{0111100}^{\texttt{m1}},\,
    j_{0111010}^{\texttt{m2}},\,j_{0111020}^{\texttt{m2}},\,j_{1111000}^{\texttt{m2}}\right)
\end{gather*}
and construct the differential systems with respect to $m_e^2$ and $s$:
\begin{equation}
    \partial_{m_e^2}\boldsymbol{j}={M}_{m_e^2}\boldsymbol{j}\,, \qquad \partial_{s}\boldsymbol{j}=M_{s}\boldsymbol{j}\,.
    \label{eq:MmeMs}
\end{equation}

Since the point $m_e^2=0$ is a singular point of the first system, we search for its solution in the form of generalized power series using the Frobenius method, along the lines of Ref. \cite{Lee2018}. The result has the form
\begin{equation}
    \bs j =\tilde{U}\bs c\,,\label{eq:tU}
\end{equation}
where $\tilde{U}$ is a fundamental matrix with entries being the generalized power series in $m_e^2$, and $\bs c$ is a column of constants. 
Note that we can always redefine $\tilde{U}$ by multiplying from the right by any non-degenerate matrix independent of $m_e^2$. In particular, we can use this freedom to secure that the column of constants $\bs c$ consists of specific coefficients in the asymptotic expansion of master integrals near the point $m_e^2=0$. Namely, we secure that
\begin{gather}
    \boldsymbol{c}^\intercal=
    \left([j_1]_{m_e^0},\,[j_2]_{m_e^0},[j_5]_{m_e^0},\,[j_3]_{m_e^{2-2\e}},\,[j_4]_{m_e^{-2\e}},\,[j_6]_{m_e^{2-2\e}},\,[j_7]_{m_e^{-2\e}}\right),
\end{gather}
where $[j_k]_{m_e^\mu}$ denotes to coefficient in front of $m_e^\mu$ in small-$m_e$ asymptotic of $j_k$. The matrix $\tilde{U}$ is found in a routine way, along the lines of Ref. \cite{Lee2018}.

Note that constants $\bs c$ depend nontrivially on $s$. In order to find the differential system for  $\bs c$ with respect to $s$, we treat $\tilde{U}$ in Eq. \eqref{eq:tU} as a transformation matrix for the second system in Eq. \eqref{eq:MmeMs}. Then we obtain 
\begin{equation}
    \partial_s\bs c = \tilde{M}_s\bs c\,,
\end{equation}
where 
\begin{equation}
    \tilde{M}_s =\tilde{U}^{-1}\left[M_s \tilde{U} - \partial_s \tilde{U}\right]
\end{equation}
Three remarks are in place here. First, since $\bs c$ is independent of $m_e^2$, so is the matrix $\tilde{M}_s$. Therefore, in order to establish the exact form of $\tilde{M}_s$, it is sufficient to know only first few terms of generalized power series in $\tilde{U}$. Second, instead of inverting the matrix $ \tilde{U}$ of truncated generalized power series, we calculate $\tilde{U}^{-1}$ independently from the equation $\partial_{m_{e}^{2}}(\tilde{U}^{-1})^{\intercal}=-({M}_{m_{e}^2})^{\intercal}(\tilde{U}^{-1})^{\intercal}$. Finally, equations for constants, corresponding to different fractional powers of $m_e^2$, decouple, as they should, so that $\tilde{M}_s$ has a block diagonal form $\tilde{M}_s = \begin{pmatrix}
    \fbox{\tiny$3\times 3$}\hspace{-3.5mm}&0\\0&\fbox{\tiny$4\times 4$}
\end{pmatrix}$.

In order to find the $\e$-forms of the resulting systems, we pass to the variable $v=\sqrt{1-4/s}$. This variable has a simple physical meaning as the maximum muon velocity in c.m.f. at a given energy. We have 
\begin{gather}
    \partial_v\begin{pmatrix}c_1 \\ c_2 \\ c_3\end{pmatrix}
    = \begin{pmatrix}
        \frac{v^2 (2 \epsilon -1)+1}{v^3-v} & -\frac{3 (\epsilon -1)}{2 v} & 0
        \\
        -\frac{4}{v-v^3} & -\frac{2 \left(v^2-3\right) (\epsilon -1)}{v
            \left(v^2-1\right)} & 0 \\
        \frac{-2 v^2 \epsilon +v^2+1}{2 v-2 v^3} & \frac{3 (\epsilon -1)}{4 v}
        & \frac{2 v}{v^2-1} \\
    \end{pmatrix}\begin{pmatrix}c_1 \\ c_2 \\ c_3\end{pmatrix}\,,\\
    \partial_v\begin{pmatrix}c_4 \\ c_5 \\ c_6\\c_7\end{pmatrix}
    = 	\begin{pmatrix}
        \frac{-3 v^2+2 \epsilon -1}{v \left(v^2-1\right)} & 0 & 0 & 0 \\
        \frac{1}{8} (v^2-1) x  (\epsilon -1) & \frac{2 v}{(v^2-1)} & 0 &
        0 \\
        \frac{(v^2-1)^3 (2 \epsilon -1)}{64 v} & 0 & \frac{2 v}{(v^2-1)} & 0 \\
        \frac{1}{8} \left(v^3-v\right) (\epsilon -1) & 0 & 0 & -\frac{2 (2 v
            \epsilon -v)}{v^2-1} \\
    \end{pmatrix} \begin{pmatrix}c_4 \\ c_5 \\ c_6\\c_7\end{pmatrix}\,.\label{eq:coefficient blocks}
\end{gather}

We reduce both systems to $\e$-form \cite{Henn2013} using \Libra, Ref. \cite{Libra}. The boundary conditions are fixed by evaluating the small-$v$ asymptotics of $j_1$ and $j_3$:
\begin{equation}
    j_{1}\sim\frac{4\pi^{2}\Gamma(2-2\e)}{\Gamma\left(\frac{3}{2}-\epsilon\right)\Gamma(\frac{7}{2}-3\e)}\times v^{5-6\epsilon}m_e^0,\quad
    j_{3}\sim\frac{16\pi^{3/2}\Gamma(\epsilon-1)}{\Gamma\left(\frac{3}{2}-\epsilon\right)}\times v^{1-2\e}m_{e}^{2-2\epsilon}.
\end{equation}
Using these boundary conditions, we obtain all $\bs c$ in terms of harmonic polylogarithms. Then we substitute  $\bs c$ in Eq. \eqref{eq:tU} and obtain a sufficient number of terms in the generalized power series representation for master integrals $\bs j$. So, our results for master integrals have the form of generalized power series in $m_e^2$ (truncated at some order), whose coefficients are series in $\e$ expressed via polylogarithms depending on $v$.

\subsection{Results}
Expressing the cross section via master integrals and substituting our results for the latter, 
we obtain the total cross section.
\begin{multline}
    \label{eq:cs egemumu}
    \sigma_{\egemumu}=\frac{\alpha^{3}}{s}\Bigg\{\ln\left(\frac{(s-4)^{2}s}{m_{e}^{2}}\right)\bigg(\frac{\left(7s^{2}-66s+50\right)v}{9s}+\frac{4\left(s^{2}+1\right)}{3s^{2}}\ln {\frac{1+v}{1-v}}\\
    -\frac{s-2}{s}\bigg[4\mathrm{Li}_{2}\left(\frac{1-v}{2}\right)+\ln {s}\,\ln {\frac{1+v}{1-v}}\bigg]\bigg)+\frac{8(s-2)}{s}\Bigg[2\mathrm{Li}_{3}\left(\frac{2v}{1+v}\right)-3\mathrm{Li}_{3}\left(\frac{1-v}{2}\right)-4\mathrm{Li}_{3}(v)\Bigg]\\
    -\frac{1}{2}\ln {\frac{1+v}{1-v}}\Bigg[\frac{14s^{3}-107s^{2}+504s-110}{9s^{2}}+\frac{4(s+3)\ln {s}}{3s}+\frac{s-2}{s}\Big(\frac{7}{3}\ln^{2} {\frac{1+v}{1-v}}+2\pi^{2}-3\ln^{2}{s}\ \Big)\Bigg]\\-\frac{\left(436s^{2}-4111s+1302\right)v}{108s}-\frac{32\left(s^{2}+1\right)\mathrm{Li}_{2}(v)}{3s^{2}}-\frac{8(s+3)\mathrm{Li}_{2}\left(\frac{1-v}{2}\right)}{3s}-\left(v\to-v\right)\Bigg\}.
\end{multline}

Diagrams from Fig. \ref{fig:egemumudiagrams} can be divided into two groups according to C-parity of the muon pair. We present for the reference the contribution of $C$-odd diagrams (the first three diagrams in Fig. \ref{fig:egemumudiagrams}) separately:
\begin{multline}
    \sigma^{odd}_{\egemumu}=\frac{\alpha^{3}}{s}\Bigg\{\ln\left(\frac{(s-4)^{2}s}{m_{e}^{2}}\right)\Biggl(\frac{s^{2}-12}{3s^{2}}\ln {\frac{1+v}{1-v}}-\frac{(8s-26)v}{9s}\Biggr)\\+\frac{(17s-154)v}{36s}
    +\frac{1}{2} \ln {\frac{1+v}{1-v}}\,\left(\frac{\left(35s^{2}-144s+198\right)}{9s^{2}}-\frac{4\left(s^{2}-12\right)\ln{s}}{3s^{2}}\right)\\
    -\frac{8\left(s^{2}-12\right)}{3s^{2}}\left[\mathrm{Li}_{2}\left(v\right)+\mathrm{Li}_{2}\left(\frac{1-v}{2}\right)\right]-(v\to-v)\Bigg\}.
\end{multline}

\subsection{Cross-section near the threshold}

When the muon kinetic energy $\sim v^2$ is comparable with $m_e$ the derived formulae are inapplicable. Although this narrow region might be not very relevant for the experiment, let us derive the appropriate expression for the cross section for the sake of completeness. For this purpose we introduce the variable $\tau$ via $\sqrt{s}=2+(1+\tau)m_{e}$. We obtain the differential systems with respect to $\tau$ and $m_e$. Then we search for master integrals  as a generalized power series in $m_e^2$, but this time, at fixed $\tau$. We need to fix two nonzero coefficients $\bs c^\intercal = \left([j_1]_{m_e^{7/2-3\e}},\,[j_1]_{m_e^{5/2-3\e}}\right)\,$. After expressing the cross-section via $\boldsymbol{c}$ we have established that $\e^0$ terms of $\boldsymbol{c}$ are sufficient for our purpose. Then we can put $\e=0$, and obtain the system $\partial_\tau \boldsymbol{c} = M\boldsymbol{c}$ with
\begin{equation}
    M=\frac{1}{3\tau(\tau+2)}\left(\begin{array}{cc}
        21(\tau+1) & \frac{9}{16}\left(11\tau^2+22\tau+12\right)\\
        -28 & -9(\tau+1)
    \end{array}\right).
\end{equation}

The solution of this system can be expressed via complete elliptic integrals $\mathrm{K}(-\tau/2)$ and  $\mathrm{E}(-\tau/2)$. We finally get

\begin{equation}
    \sigma^{threshold}_{\egemumu}=\frac{\sqrt{2}\alpha^{3}}{3}\left(m_e\right)^{3/2}\left(-8(\tau+1) \mathrm E\left(-\frac{\tau}{2}\right)+(\tau+2)(3\tau+4)\mathrm K\left(-\frac{\tau}{2}\right)\right)\label{eq:cs egemumu threshold}.
\end{equation}
We remind here that  $\tau=\frac{\sqrt{s}-2-m_e}{m_e}$ and that the formula above is valid when $0<\sqrt{s}-2-m_e\ll 1$, while Eq. \eqref{eq:cs egemumu} is valid when $m_e\ll \sqrt{s}-2-m_e$. The two formulae agree with each other in the overlapping region  $m_e\ll \sqrt{s}-2-m_e \ll 1$.

\begin{figure}
    \centering
    \includegraphics[width=0.6\linewidth]{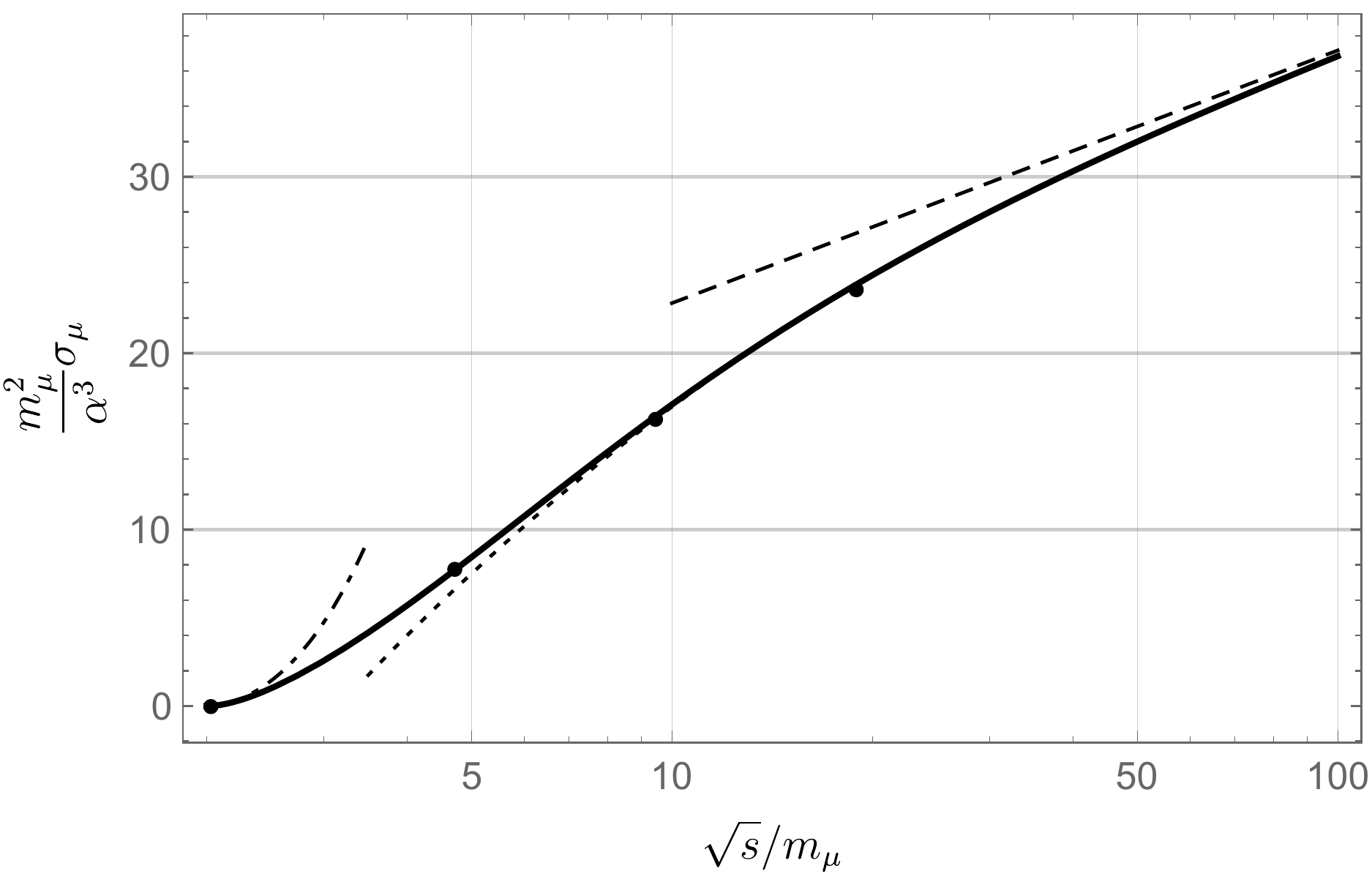}
    \caption{Exact cross-section of the $\egemumu$ (solid curve) and  asymptotics. Black dashed curve --- leading asymptotic, dotted curve --- Eq. \eqref{eq:cs egemumu high energy asymp}, and dash-dotted curve --- the cross-section near threshold, Eq. \eqref{eq:cs egemumu threshold}. Points represent the data obtained by numerical Monte-Carlo integration of the differential cross section using \texttt{Cuba} library.}
    \label{fig:cs_egemumu}
\end{figure}

In Fig. \ref{fig:cs_egemumu} we present our exact result \eqref{eq:cs egemumu} and the found threshold and high-energy asymptotics. As a cross-check, we present on the same figure also a few points obtained by numerical integration of the differential cross section using the \texttt{Cuba} library \cite{hahn2016concurrent}.

\subsection{High-energy asymptotics}

For the high-energy asymptotics of the total cross-section of $\egemumu$  we have
\begin{multline}
    \sigma_{\egemumu}=\alpha^3\,\Bigg(\left[\frac{28}{9}\ln\left(\frac{s}{m_{e}}\right)-\frac{218}{27}\right]+\frac{4}{3s}\Bigg\{\pi^2\ln\left(\frac{s^2}{m_e}\right)-\ln^2{s} \ln\left(\frac{s^2}{m_e^3}\right)\\
    +\ln{s} \ln \left(\frac{s^5}{m_e^4}\right)+\frac{80}{3}\ln{m_e} -\frac{373}{12}\ln{s}-12\zeta_3-\frac{5}{3}\pi^{2}+\frac{1493}{24}\Bigg\}+O\left({s^{-2}}\right)\Bigg).\label{eq:cs egemumu high energy asymp}
\end{multline}

The leading term of this asymptotics has been derived in Ref. \cite{Baier} using the equivalent photon approximation. It coincides with the leading term of our result.

\subsection{Total cross-section of $e^-\gamma\to e^- \pi^+\pi^-$}
For the sake of completeness, let us present also the total cross section of $e^-\gamma\to e^-  \pi^+\pi^-$, where we consider $\pi$-meson as a pointlike scalar particle. We  have
\begin{multline}\label{eq:cs egepipi}
    \sigma_{e^-\gamma\to e^-\pi^+\pi^-}=\frac{\alpha^{3}}{s^2}\Biggl\{\ln\left(\frac{(s-4)^{2}s}{m_{e}^{2}}\right)\Biggl(\frac{\left(4s^{2}+87s-118\right)v}{36}-4\mathrm{Li}_{2}\left(\frac{1-v}{2}\right)\\
    -\ln {\frac{1+v}{1-v}}\Biggl[\frac{11s^{2}+20}{12s}+\ln{s}\Biggr]\Biggr)
    +\ln{\frac{1+v}{1-v}}\Biggl(\frac{5s^{2}+6s+12}{6s}\ln{s}-\frac{8s^{3}-155s^{2}-756s-50}{72s}\\
    -\frac{7}{6} \ln^2 {\frac{1+v}{1-v}}-\pi^{2}+\frac{3}{2}\ln^{2}{s}\Biggr)
    +\frac{2\left(11s^{2}+20\right)\mathrm{Li}_{2}(v)}{3s}
    +\frac{2 \left(5 s^2+6 s+12\right) \mathrm{Li}_2\left(\frac{1-v}{2}\right)}{3s}\\
    -24\mathrm{Li}_{3}\left(\frac{1-v}{2}\right)
    -32\mathrm{Li}_{3}(v)
    +16\mathrm{Li}_{3}\left(\frac{2v}{1+v}\right)
    -\frac{\left(208s^{2}+7547s-4278\right)v}{432}-(v\to-v)\Biggr\}.
\end{multline}
Note that the leading high-energy asymptotics of this formula, $\alpha^{3}\left[\frac{4}{9}\ln(s/m_e)-\frac{26}{27}\right]$, perfectly agrees with the result of Ref. \cite{Baier}.

\section{Total cross section of the process $e^-\gamma \to e^-\gamma\gamma$}

In this Section we present the calculation of the cross section of the process \egegg (double Compton scattering) at arbitrary energies. To avoid the infrared divergences, we restrict the integration region by the condition $\omega_{2,3}>\omega_0$, where $\omega_{2,3}$ are the energies of the outgoing photons, and $\omega_0$ is a small cut-off parameter. This restriction is frame-dependent, and we consider two physically relevant frames: the center-of-mass frame and the initial electron rest frame. From the technical point of view, we proceed as follows. Using dimensional regularization, we first calculate the total cross section, which contains the term $\propto \frac{1}{\epsilon}$. Then we calculate separately the contribution of soft-photon region and subtract it from the total cross section to obtain the physically observed cross section $\sigma_{\egegg}(\omega_0)$.
\subsection{Calculation of the total cross section in dimensional regularization}
\begin{figure}
    \centering
    \includegraphics[width=0.99\linewidth]{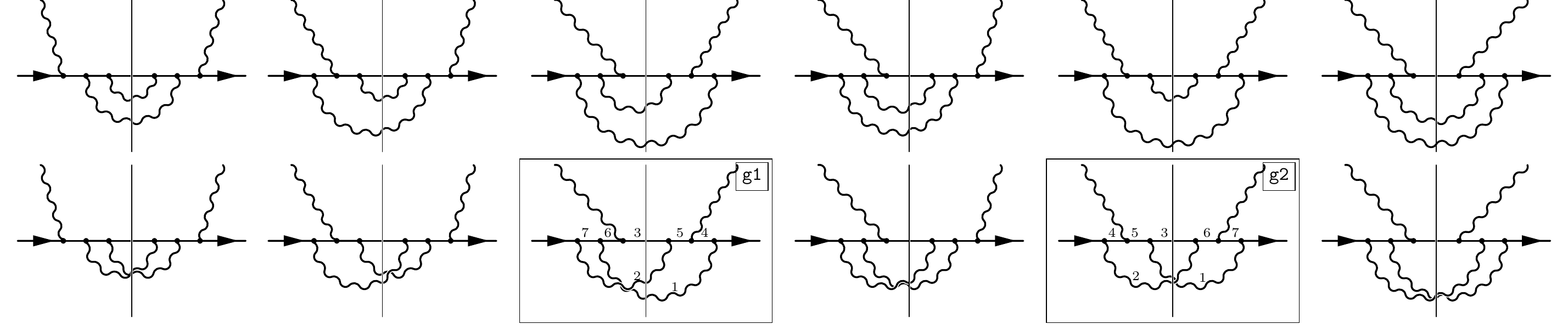}
    \caption{Diagrams contributing to the total cross section of the \egegg process. 
    }
    \label{fig:egeggudiagrams}
\end{figure}

Using the optical theorem \eqref{intro:ot}, the total cross section \egegg can be expressed in terms of cut diagrams shown in Fig. \ref{fig:egeggudiagrams}.
We set up two \LiteRed bases \texttt{g1} and \texttt{g2} containing propagators of boxed diagrams in Fig. \ref{fig:egeggudiagrams}.
We find $14$ master integrals,
\begin{gather*}
    j_{1110000}^{\texttt{g1}},\thinspace
    j_{111000-1}^{\texttt{g1}},\thinspace 
    j_{1110001}^{\texttt{g1}},\thinspace
    j_{1110002}^{\texttt{g1}},\thinspace
    j_{1110011}^{\texttt{g1}},\thinspace
    j_{1110110}^{\texttt{g1}},\thinspace
    j_{1111001}^{\texttt{g1}},\\
    j_{1110120}^{\texttt{g1}},\thinspace
    j_{1111002}^{\texttt{g1}},\thinspace
    j_{1110111}^{\texttt{g1}},\thinspace
    j_{1110112}^{\texttt{g1}},\thinspace
    j_{1110110}^{\texttt{g2}},\thinspace
    j_{1110111}^{\texttt{g2}},\thinspace
    j_{1111111}^{\texttt{g2}}.
\end{gather*}
Introducing the column-vector $\boldsymbol{j}=\left(j_1,..,j_{14}\right)^{\intercal}= (j_{1110000}^{\texttt{g1}},..,j_{1111111}^{\texttt{g2}})^{\intercal}$ we obtain the differential system
\begin{equation}
    \partial_{s}\boldsymbol{j}=M\left(s,\epsilon\right)\boldsymbol{j},
    \label{eq:egeggde} 
\end{equation}
where $M$ is the matrix, rationally depending on $s$ and $\epsilon$.  Introducing the new variable $y=\sqrt{\frac{s-1}{s+3}}$, we reduce the system \eqref{eq:egeggde} to $\epsilon$-form using \texttt{Libra} \cite{Libra}:
\begin{equation}
    \partial_{y}\boldsymbol{J}\left(y\right)=\epsilon\left[\frac{1}{y}S_{0}+\frac{1}{y-1}S_{1}+\frac{1}{y+1}S_{2}+\frac{2y}{y^2+1/3}S_{3}\right]\boldsymbol{J}\left(y\right),
    \label{eq:egegdey} 
\end{equation}
where $S_i$ are some constant matrices. The canonical basis $\boldsymbol{J}$ is connected with $\boldsymbol{j}$ by rational transformation $\boldsymbol{j}=T\left(y,\epsilon\right)\boldsymbol{J}$. 
The general solution of the system \eqref{eq:egegdey}, $\boldsymbol{J}=\sum_{n}\epsilon^{n}\boldsymbol{J}^{\left(n\right)}$, can be easily written in terms of generalized polylogarithms. We fix the boundary conditions by calculating the coefficients in the asymptotic expansion of the master integrals at the threshold. The only nontrivial coefficient of the leading threshold asymptotic is 
\begin{equation}
    j_{2}\sim-y^{8-8\epsilon}\frac{2^{5-8\epsilon}\pi\Gamma^{2}\left(1-\epsilon\right)}{\Gamma\left(4-4\epsilon\right)}.
    \label{eq:egeggasym} 
\end{equation}
Using this boundary condition, we find an expression for the master integrals and get the total Born cross sections \egegg in the form
\begin{equation}
    \sigma_{\egegg}=\frac{\sigma_{1}}{\epsilon}+\sigma_{2}.
    \label{eq:egeggform} 
\end{equation}
where $\sigma_{1},\sigma_{2}$ can be expressed via generalized polylogarithms with letters $0,\pm1, \pm\frac{i}{\sqrt{3}}$. In this expression we omit the terms suppressed by $\epsilon$. The $\frac{1}{\epsilon}$ term is quite anticipated, it is due to the contribution of soft-photon region. In order to get rid of this term, we have to subtract the contribution related to soft photons.
\subsection{Soft-photon contribution to \egegg}
In a soft-photon approximation the differential cross section factorizes as 
\begin{equation}
    d\sigma^{soft}_{\egegg}=d\sigma_{e^-\gamma\rightarrow e^-\gamma}\cdot dW_{\gamma}, \label{eq:egeggsoft}
\end{equation}
where $dW_{\gamma}$ is a soft-photon emission probability
\begin{equation}
    dW_{\gamma}=-\frac{d^{d-1}\boldsymbol{k}_{3}}{\left(2\pi\right)^{d-1}2\omega_{3}}e^{2}\left( \frac{p_{1}}{p_{1}\cdot k_{3}}-\frac{p_{2}}{p_{2}\cdot k_{3}}\right)^{2}.\label{eq:egeggdW}
\end{equation}
Here $k_3$ is the four-momentum  of the soft photon. The contribution to the total cross section can be written as
\begin{equation}
    \sigma^{soft}_{\egegg}=\int d\sigma_{e^-\gamma\rightarrow e^-\gamma}dW_{\gamma}\,\theta\left(\omega_{0}-\omega_{3}\right).\label{eq:egeggcss}
\end{equation}
Here $\omega_{0}$ is the maximal energy of soft photon.  As we have already mentioned, the infrared cutoff introduces the frame dependence. We calculate the cross section in the center-of-mass frame (cmf) and in the rest frame of the initial electron (rf). Below we present some details of the calculation for the electron rest frame.

The soft-photon contribution \eqref{eq:egeggcss} can formally be written as a two-loop integral with cut propagators. Inserting a identity  $1=\int d\omega\delta\left(\omega-\omega_{3}\right)$  into \eqref{eq:egeggcss} and rescaling $k_{3}\rightarrow \omega k_{3}$ we get
\begin{equation}
    \sigma^{soft}_{\egegg}=-\frac{\omega^{-2\epsilon}_{0}}{2\epsilon}\int d\sigma_{e^-\gamma\rightarrow e^-\gamma}\frac{d^{d}k_{3}}{\left(2\pi\right)^{d-1}}\delta\left(k_{3}^{2}\right)\left(\frac{p_{1}}{p_{1}\cdot k_{3}}-\frac{p_{2}}{p_{2}\cdot k_{3}}\right)^{2}\delta\left(p_{1}\cdot k_{3}-1\right). \label{eq:egeggsigma}
\end{equation}
We calculate the integral on the right side using the differential equations method. This integral should be considered up to the $O(\epsilon^{1})$ term.

To determine the basis for IBP  reduction, we include propagators from single Compton scattering, scalar products in denominators in \eqref{eq:egeggcss}, and cut propagators from $\delta$-functions. We obtain scalar integral
\begin{equation}
    j_{n_{1},..,n_{7}}^{soft}=\int\frac{d^{d}k_{2}d^{d}k_{3}}{\left(p_{2}\cdot k_{3}\right)^{n_{5}}}\frac{\left(1-p_{1}\cdot k_{3}\right)^{-n_{4}}\left(k_{1}\cdot k_{3}\right)^{-n_{7}}}{\left(k_{2}^{2}\right)^{n_{1}}\left(k_{3}^{2}\right)^{n_{2}}\left(p_{2}^{2}-1\right)^{n_{3}}\left(\left(p_{1}-k_{2}\right)^{2}-1\right)^{n_{6}}},\label{eq:egeggscint}
\end{equation}
where $p_2 = p_1 + k_1 - k_2$. We imply that the first four propagators in this integral are cut, and that the last one is the irreducible numerator. Using \LiteRed package we reveals the following master integrals
\begin{equation}
    j_{1111000}^{soft},\quad j_{1111010}^{soft},\quad j_{1111100}^{soft},\quad
    j_{1111200}^{soft},\quad j_{111100-1}^{soft},\quad j_{1111110}^{soft}.\label{eq:egeggj}
\end{equation}
Using \Libra we find the canonical basis $\boldsymbol{J}$ related to $\boldsymbol{j}^{soft}$ by
\begin{gather}
    j_{1111000}^{soft}=\left(s-1\right)J_{1},\qquad j_{1111010}^{soft}=\frac{s(2\epsilon-1)}{3(s-1)\epsilon}J_{2},\quad j_{1111100}^{soft}=\frac{2(2\epsilon-1)^{2}}{9(s-1)\epsilon^{2}}J_{4}\nonumber\\
    j_{1111200}^{soft}=\frac{2(s+1)^{2}(2\epsilon-1)^{2}}{9(s-1)s^{2}}J_{4}-\frac{2(s+1)(2\epsilon-1)^{2}}{9s^{2}}J_{3}-\frac{2(2\epsilon-1)^{2}}{s^{2}}J_{1},\nonumber\\
    j_{111100-1}^{soft}=\frac{s(2\epsilon-1)}{3\epsilon}J_{5}-\frac{2s(2\epsilon-1)^{2}}{9(s-1)\epsilon^{2}}J_{4},\quad j_{1111110}^{soft}=\frac{16(2\epsilon-1)^{2}}{9y(s-1)(s+3)\epsilon^{2}}J_{6}.
    \label{eq:egeggT}
\end{gather}
The differential equations for $\boldsymbol{J}$ have the form
\begin{gather}
    J_{1}'=\frac{1-3s}{(s-1)s}\epsilon J_{1},\quad J_{2}'=\frac{3}{s}\epsilon J_{1}-\frac{2}{s}\epsilon J_{2},\quad J_{3}'=\frac{9(3-s)}{2(s-1)s}\epsilon J_{1}-\frac{2}{s}\epsilon J_{3},\nonumber\\
    J_{4}'=-\frac{9}{2s}\epsilon J_{1}-\frac{1}{s}\epsilon J_{3}-\frac{s-3}{(s-1)s}\epsilon J_{4}, \quad
    J_{5}'=-\frac{3}{s}\epsilon J_{1}-\frac{2}{s}\epsilon J_{5},\nonumber\\
    J_{6}'=\frac{9\epsilon J_{1}}{8y(s+3)}-\frac{3\epsilon J_{2}}{8y(s+3)}+\frac{\epsilon J_{3}}{4y(s+3)}-\frac{2(s-3)\epsilon J_{6}}{(s-1)s(s+3)}.\label{eq:egeggdes}
\end{gather}
To fix the boundary conditions it suffices to calculate the leading asymptotics of $j_{1111000}^{soft}$,
\begin{equation}
    j_{1111000}^{soft}\sim-y^{2-4\epsilon}\frac{2^{3-2\epsilon}\pi^{2}}{\Gamma^{2}\left(\frac{3}{2}-\epsilon\right)}.
\end{equation}

Finally, we obtain the contribution of soft-photons in the form
\begin{equation}
    \sigma^{soft}_{\egegg}=\frac{\sigma_{1}}{\epsilon}+\sigma_{3}\cdot\ln{\omega_{0}}+\sigma_{4}.\label{eq:egeggcsf}
\end{equation}
Note that $\tfrac1{\e}$ term in this formula appears to be the same as in Eq. \eqref{eq:egeggform}, as it should be. Therefore, in the difference the terms, containing $\tfrac{1}{\epsilon}$, vanish.
\subsection{Results}
The total Born cross section of the process \egegg, integrated over the kinematic region, in which $\omega_{2,3}>\omega_{0}$, in the electron rest frame
\begin{multline}
    \sigma_{\egegg}^{rf}\left(\omega_{0}\right)=\alpha^{3}\,\Bigg\{\frac{\left(3s^{5}+3s^{4}+62s^{3}+222s^{2}+15s-465\right)yA_{2}\left(y\right)}{(s-1)^{4}(s+3)}+\frac{20B_{3}\left(y\right)}{(s-1)^{2}}\\
    +\frac{12\left(s^{4}-3s^{3}-27s^{2}-33s+14\right)yC_{3}\left(y\right)}{(s-1)^{4}(s+3)}+\frac{6(s+1)\left(s^{3}-7s^{2}-29s-13\right)yD_{3}\left(y\right)}{(s-1)^{4}(s+3)}\\
    +\frac{6\left(s^{2}-4s-9\right)A_{3}\left(y\right)}{(s-1)^{3}}+\frac{6\left(s^{4}-2s^{3}-24s^{2}-30s+23\right)yE_{3}\left(y\right)}{(s-1)^{4}(s+3)}+\frac{51s^{3}+2629s^{2}-459s+99}{24(s-1)^{2}s^{2}}\\
    +\frac{\left(5s^{5}-295s^{4}-1061s^{3}-161s^{2}+108s-36\right)\ln{s}}{4(s-1)^{3}s^{2}(s+3)}-\frac{\left(4s^{5}+s^{4}-36s^{3}-60s^{2}-28s-9\right)\ln^{2}{s}}{2(s-1)^{4}(s+3)}\\
    -\frac{\left(s^{5}+4s^{4}-44s^{3}-98s^{2}+35s+6\right)\text{Li}_{2}\left(1-s\right)}{(s-1)^{3}s(s+3)}+\frac{2(5s+1)\ln {s}\text{Li}_{2}\left(1-s\right)}{(s-1)^{3}}\\
    +\ln\left(\frac{s-1}{2\omega_{0}}\right)\bigg(\frac{12\left(s^{4}-2s^{3}-24s^{2}-30s+7\right)yA_{2}\left(y\right)}{(s-1)^{4}(s+3)}-\frac{5s^{3}+45s^{2}-5s+3}{(s-1)^{2}s^{2}}\\
    -\frac{2\left(s^{5}-27s^{4}-86s^{3}-34s^{2}+5s-3\right)\ln{s}}{(s-1)^{3}s^{2}(s+3)}-
    \frac{2\left(s^{2}-6s-1\right)\ln^{2}{s}}{(s-1)^{3}}\bigg)\Bigg\}. \label{eq:egeggcsrf}
\end{multline}
For center-of-mass frame we have
\begin{multline}
    \sigma_{\egegg}^{cmf}=\sigma_{\egegg}^{rf}+\alpha^{3}\,\Bigg\{\frac{6\left(s^{4}-6s^{3}-28s^{2}-26s+11\right)yA_{2}\left(y\right)}{(s-1)^{4}(s+3)}\\
    -\frac{24\left(s^{4}-2s^{3}-24s^{2}-30s+7\right)yD_{3}\left(y\right)}{(s-1)^{4}(s+3)}-\frac{\left(7s^{6}-16s^{5}-67s^{4}-8s^{3}-39s^{2}-8s+3\right)\ln^{2}{s}}{2(s-1)^{4}s^{2}(s+3)}\\	-\ln{s}\left(\frac{6\left(s^{4}-2s^{3}-24s^{2}-30s+7\right)yA_{2}\left(y\right)}{(s-1)^{4}(s+3)}-\frac{5s^{5}+29s^{4}+24s^{3}-16s^{2}-29s+3}{(s-1)^{3}s^{2}(s+3)}\right)\\
    -\frac{12\left(s^{4}-2s^{3}-24s^{2}-30s+7\right)yC_{3}\left(y\right)}{(s-1)^{4}(s+3)}-\frac{5s^{3}+9s^{2}-5s-1}{(s-1)^{2}s^{2}}\Bigg\}.
    \label{eq:egeggcscmf}
\end{multline}
Here $A_{2},A_{3},B_{3},C_{3},D_{3},E_{3}$ are functions which can be expressed via classical polylogarithms. They are defined as follows
\begin{gather}
    A_{2}\left(y\right)=  \Re\left[\text{Li}_{2}\left(e^{\frac{i\pi}{3}}q\right)-\text{Li}_{2}\left(\frac{e^{\frac{i\pi}{3}}}{q}\right)\right],\quad A_{3}\left(y\right)=\Re\left[\text{Li}_{3}\left(e^{\frac{i\pi}{3}}q\right)+\text{Li}_{3}\left(\frac{e^{\frac{i\pi}{3}}}{q}\right)\right]-\frac{2}{3}\zeta_{3},\nonumber \\
    B_{3}\left(y\right)=  \frac{\pi^{2}}{6}\ln{s}+\frac{1}{2}\ln\left(s-1\right)\ln^{2}s-\frac{1}{6}\ln^{3}s+\text{Li}_{2}\left(1-s\right)\ln s+\text{Li}_{3}\left(\frac{1}{s}\right)-\zeta_{3},\nonumber \\
    C_{3}\left(y\right)=  \int_{1}^{s}\frac{\ln\left(x\right)\ln\left(\frac{s+3}{x+3}\right)}{\sqrt{\left(x-1\right)\left(x+3\right)}}dx,\ D_{3}\left(y\right)=-\int_{1}^{s}\frac{\ln^{2}\left(x\right)}{2\sqrt{\left(x-1\right)\left(x+3\right)}}dx,\nonumber \\
    E_{3}\left(y\right)=  \int_{1}^{s}\frac{\text{Li}_{2}\left(1-x\right)}{\sqrt{\left(x-1\right)\left(x+3\right)}}dx,
\end{gather}
where $q=\frac{1-y}{1+y}$, $y=\sqrt{\frac{s-1}{s+3}}$. The integrals in $C_3$, $D_3$ and $E_3$ can be taken in terms of $\text{Li}_{3}$ and similar functions; the corresponding expressions are quite lengthy and given in the ancillary file \texttt{CDEfunctions.m}.

\begin{figure}
    \centering
    \includegraphics[width=0.6\linewidth]{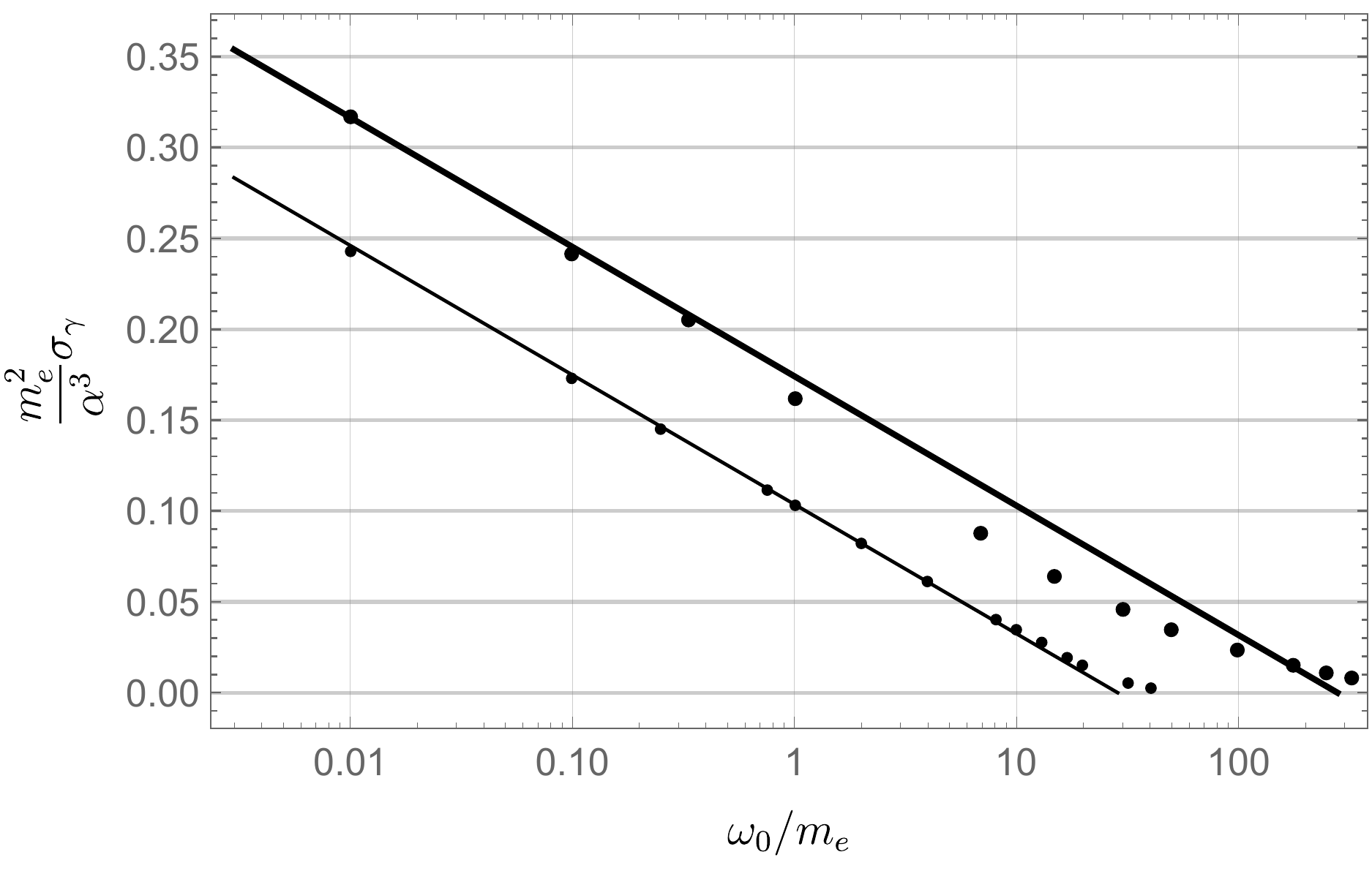}
    \caption{Dependence of the exact cross section on the infrared cutoff parameter $\omega_0$ at $\sqrt{s}=100$. The upper thick line shows the cross section in electron rest frame, the lower thin line --- the cross section in center-of-mass frame, points correspond to  numerical integration of the differential cross section with \texttt{Cuba} library.}
    \label{fig:egeggCuba}
\end{figure}

Let us discuss the limits of applicability of the soft photon approximation used in the present work. For the low-energy region it follows from \eqref{eq:egegggould} that we should require $\frac{s-1}{2\omega_{0}}\gg1$, so that the logarithm dominates. For the high-energy region, the requirements for $\omega_{0}$ is essentially different for the two reference frames we consider. In the center-of-mass frame the infrared cut-off parameter should be small compared to $\sqrt{s}$. Thus, it is natural to expect that in the electron rest frame the soft photon approximation breaks down already at $\omega_0\gtrsim m_e=1$. This is because the photon with energy $\sim m_e$ in the electron rest frame can have energy $\sim \sqrt{s}$ when boosted to the center-of-mass frame. This is demonstrated in Fig. \ref{fig:egeggCuba}, where the results of the present paper are compared with those of numerical integration using the \texttt{Cuba} library \cite{hahn2016concurrent}.
\begin{figure}
    \centering
    \includegraphics[width=0.6\linewidth]{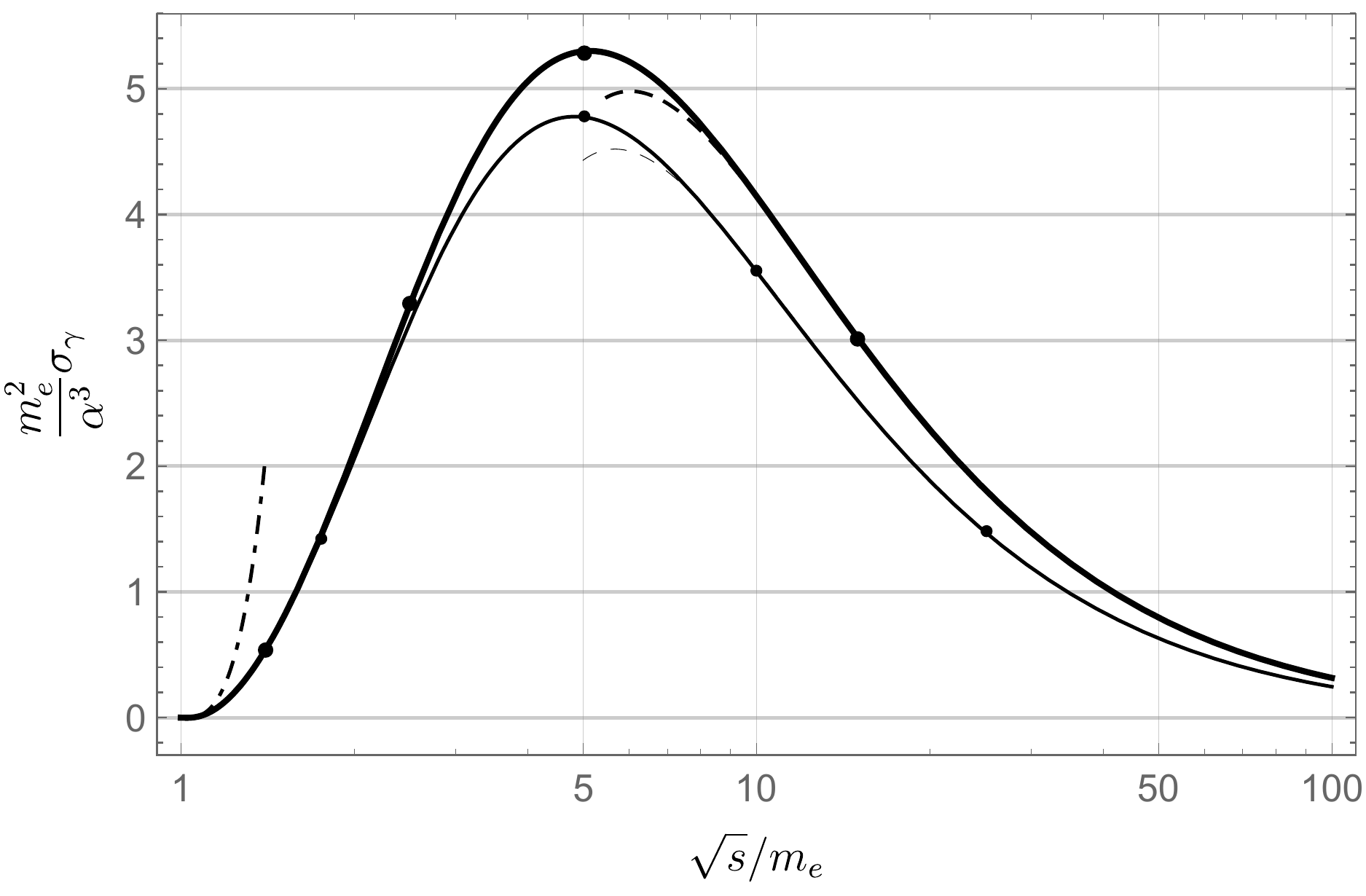}
    \caption{Exact cross section as a function of $\sqrt{s}/m_e$ at $\omega_{0}=0.01m_{e}$. The upper thick line --- the cross section in electron rest frame, the lower thin line --- in center-of-mass frame, dashed lines --- the high-energy asymptotics \eqref{eq:egeggasymrf},\eqref{eq:egeggasymcmf}, dash-dotted line --- the threshold asymptotics \eqref{eq:egegggould}, points --- \texttt{Cuba} numerical integration of the differential cross section.}
    \label{fig:egegg2cs}
\end{figure}

Fig. \ref{fig:egegg2cs} shows the dependence of the cross section on $s$ at $\omega_{0}=0.01$. On the same figure we have shown a few points obtained by numerical integration of the differential cross section using \texttt{Cuba} library.
We have also performed the comparison of the obtained results with known numerical and/or approximate results. First, the numerical result of Ref. \cite{ram1971calculation} disagrees with our result by a factor of $2$. The discrepancy is possibly due to the overlooked Bose symmetry factor $\frac{1}{2!}$ in Ref. \cite{ram1971calculation}. Once this factor is recovered, we find perfect agreement. 
Numerical results of Ref. \cite{lotstedt2013theoretical} agree with our results up to $s\approx 30m_e^2$. The discrepancy for higher energies is quite expected due to inapplicability of the soft-photon approximation for the parameter $\omega_0=\frac{\omega_1}{50}=\frac{s- m_e^2}{100 m_e}$ chosen in Ref. \cite{lotstedt2013theoretical}.
\subsection{Asymptotics}
Using the expressions for the exact cross sections \eqref{eq:egeggcsrf} and \eqref{eq:egeggcscmf} via polylogarithms, it is easy to calculate both the threshold asymptotics and the high-energy asymptotics.

Threshold asymptotics 
\begin{equation}
    \sigma_{\egegg}\left(\omega_{0}\right)\stackrel{s\rightarrow1}{=}\frac{8\alpha^{3}}{9}\left( s-1\right) ^{2}\left(\ln\left(\frac{s-1}{2\omega_{0}}\right)-\frac{167}{120}\right)+O\left(s-1\right)^{3}
    \label{eq:egegggould}	
\end{equation}
coincides with the non-relativistic asymptotic in Ref. \cite{gould1984cross}. We remind that this formula implies that $\omega_{0}\ll s-1\ll 1$. Note that this asymptotics holds both for electron rest frame and for the center-of-mass frame.

The high-energy asymptotics has the form
\begin{multline}
    \sigma_{\egegg}^{rf}\left(\omega_{0}\right)\stackrel{s\rightarrow\infty}{=}\frac{\alpha^{3}}{s}\Bigg\{\ln\left(\frac{\sqrt{s}}{2\omega_{0}}\right)\left(4\ln^{2}{s}-2\ln{s}-5-\frac{2\pi^{2}}{3}\right)\\
    +\ln^{3}{s}-\frac{3\ln^{2}{s}}{2}-\ln{s}\left(\frac{5\pi^{2}}{3}-\frac{3}{4}\right)+10\zeta_3+\frac{2\pi^{2}}{3}+\frac{33}{8}+O\left(s^{-1}\right)\Bigg\}
    \label{eq:egeggasymrf}
\end{multline}
in the electron rest frame, and 
\begin{multline}
    \sigma_{\egegg}^{cmf}\left(\omega_0\right)\stackrel{s\rightarrow\infty}{=}
    \frac{\alpha^{3}}{s}
    \Bigg\{
    \ln\left(\frac{\sqrt{s}}{2\omega_{0}}\right)\left(4\ln^{2}{s}-2\ln{s}-5-\frac{2\pi^{2}}{3}\right)\\
    -2\ln^{2}{s}-\ln{s}\left(\frac{2\pi^{2}}{3}-\frac{23}{4}\right)+2\zeta_3+\frac{\pi^{2}}{3}-\frac{7}{8}+O\left(s^{-1}\right)\Bigg\}
    \label{eq:egeggasymcmf}
\end{multline}
in center-of-mass frame.
We note that the high-energy asymptotics of the double Compton scattering is obtained for the first time here to the best of our knowledge.

\section{Total cross section of the process \egeee.}

The total cross section of the process \egeee is not expressible via polylogarithms. This is easy to understand because among the relevant master integrals there is an equal mass two-loop sunrise integrals which is a classical example of non-polylogarithmic integrals. Therefore, we will rely here on the approach of Ref. \cite{Lee2019}. We construct an $\e$-regular basis sufficient for the calculation of the cross section\footnote{Note that the cross section has no IR divergences.}.

\begin{figure}
    \centering
    \includegraphics[width=0.83\linewidth]{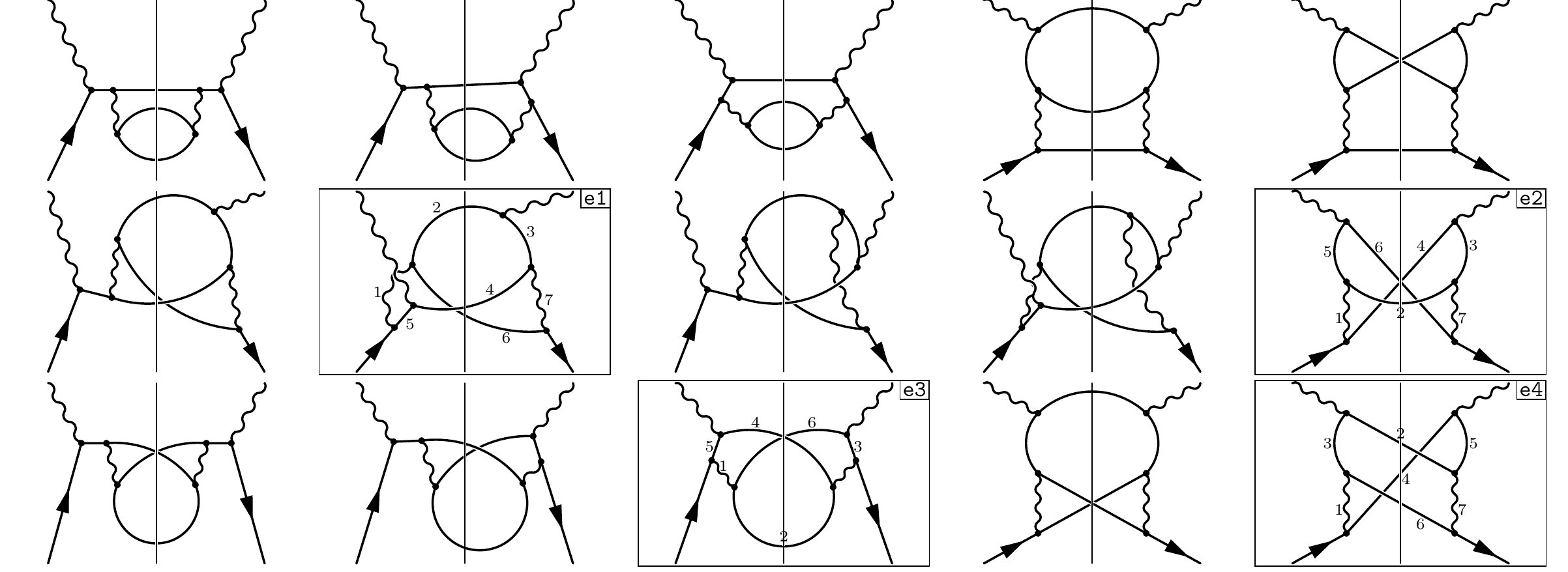}
    \caption{Diagrams contributing to the total cross section of the \egeee process.}
    \label{fig:egeeediagrams}
\end{figure}

The cut diagrams which contribute to the cross section are shown in Fig. \ref{fig:egeeediagrams}. We define 4 \LiteRed bases \texttt{e1}, \texttt{e2}, \texttt{e3}, \texttt{e4}, corresponding to the denominators of the framed diagrams in the same figure. We find 18 master integrals which we choose as follows:
\begin{gather*}
    j_{0101010}^{\texttt{e1}},\,j_{0101020}^{\texttt{e1}},\,j_{0101011}^{\texttt{e1}},\,j_{0101111}^{\texttt{e1}},\,j_{1101011}^{\texttt{e1}},\,j_{1101-111}^{\texttt{e1}},\,j_{1111010}^{\texttt{e1}},\,j_{1111011}^{\texttt{e1}},\,j_{1111-11}^{\texttt{e1}},\,\\
    j_{1111110}^{\texttt{e1}},\,
    j_{0101111}^{\texttt{e2}},\,j_{1101011}^{\texttt{e2}},\,j_{1101111}^{\texttt{e2}},\,j_{0111111}^{\texttt{e2}},\,j_{1111111}^{\texttt{e2}},\,j_{1101011}^{\texttt{e3}},\,j_{1101111}^{\texttt{e3}},\,j_{1111111}^{\texttt{e3}}.
\end{gather*}
Note that the integrals $j_{0101111}^{\texttt{e1}}$ and $j_{1101111}^{\texttt{e3}}$ can, in principle, be expressed via the integrals of the lower sectors, and can be replaced by, e.g.,  $j_{0101110}^{\texttt{e1}}$ and $j_{1101012}^{\texttt{e3}}$, respectively. However we find our present choice to be convenient as we empirically observe that the IBP reduction of any other integral to our set of masters does not generate inverse powers of $\e$ in the coefficients. In particular, the total cross section can be expressed as a linear combination of these master integrals with regular coefficients. The differential system for these master integrals is also regular at $\e\to 0$. Therefore, we can safely put $\e=0$ starting from this point. 

In order to further simplify the differential system, we use \Libra \cite{Libra} and pass to new master integrals $J_1,\ldots,J_{18}$ related to our previous master integrals via
\begin{gather}
    j_{0101010}^{\texttt{e1}}=(s+3) J_2-4J_1,\quad
    j_{0101020}^{\texttt{e1}}=2 J_2-2J_1,\quad
    j_{0101011}^{\texttt{e1}}=\frac{8 J_3}{s-1}-2J_1,\nonumber\\
    j_{0101111}^{\texttt{e1}}=\frac{4J_4}{s-1},\quad
    j_{1101011}^{\texttt{e1}}=-\frac{8J_4}{s-1}+\frac{4 J_6}{s-1}-\frac{4J_7}{s-1},\nonumber\\
    j_{1101-111}^{\texttt{e1}}=-\frac{2\left(-s^2+2 s+3\right) J_3}{s^2-s}-4 J_1+2J_2-\frac{2 (s-1)J_5}{s},\nonumber\\
    j_{1111010}^{\texttt{e1}}=-\frac{4J_4}{s-1}-\frac{2J_7}{s-1},\quad
    j_{1111011}^{\texttt{e1}}=\frac{4 yJ_8}{(s-1)^2},\quad
    j_{1111-11}^{\texttt{e1}}=-\frac{4J_6}{s-1}+\frac{2 J_7}{s-1}-\frac{2J_{13}}{s-1},\nonumber\\
    j_{1111110}^{\texttt{e1}}=\frac{4J_4}{(s-1)^2}+\frac{4J_{14}}{(s-1)^2},\quad
    j_{0101111}^{\texttt{e2}}=\frac{2 zJ_9}{s-1}-\frac{6J_3}{s-1},\nonumber\\
    j_{1101011}^{\texttt{e2}}=2 z J_9+\frac{8(3-s) J_3}{s-1}+2J_5+2J_{15},\quad
    j_{1101111}^{\texttt{e2}}=\frac{2 zJ_{10}}{s^2-6 s+5},\nonumber\\
    j_{0111111}^{\texttt{e2}}=-\frac{2J_{11}}{(s-1)^2},\quad
    j_{1111111}^{\texttt{e2}}=\frac{4 zJ_{12}}{(s-5) (s-1)^2},\quad
    j_{1101011}^{\texttt{e3}}=\frac{2J_5}{s}-\frac{6J_{16}}{s},\nonumber\\
    j_{1101111}^{\texttt{e3}}=-\frac{4J_6}{(s-1)^2}+\frac{2 J_7}{(s-1)^2}-\frac{2J_{13}}{(s-1)^2}-\frac{4J_{17}}{(s-1)^2},\quad
    j_{1111111}^{\texttt{e3}}=\frac{4 yJ_8}{(s-1)^3}+\frac{8 y J_{18}}{(s-1)^3}\,.
\end{gather}
Here we have introduced the notations
$y=\sqrt{\frac{s-1}{s+3}}$ and $z=\sqrt{\frac{s-5}{s-1}}$.

The differential system for $J_k$ has the form
\begin{gather}
    J_1'=\frac{8 J_1}{(s-9) (s-1)}+\frac{J_2}{s-1},\quad
    J_2'=\frac{2 (s+3) J_1+(s-9)J_2}{s(s-9)(s-1)},\quad
    J_3'=\frac{2J_1}{s-9},\quad 
    J_4'=\frac{J_3}{s-1},\nonumber\\
    J_5'=\frac{J_1}{s-1},\quad
    J_6'=\frac{J_5}{2 s},\quad
    J_7'=-\frac{3 J_3}{2s},\quad 
    J_8'=\frac{ y }{s-1}(3J_3+J_5),\quad J_9'=\frac{(7 s-15)
        z J_1}{(s-9) (s-5)},\nonumber\\
    J_{10}'=\frac{z}{s-5}\left(2J_5-9J_3+2J_{15}\right)+\frac{J_9}{s-1},\quad 
    J_{11}'=\frac{z J_9}{s-5},\quad 
    J_{12}'=\frac{3 z J_3-zJ_5+J_9-zJ_{15}}{s-5},\nonumber\\
    J_{13}'=0,\quad J_{14}'=0,\quad J_{15}'=0,\quad J_{16}'=0,\quad J_{17}'=-\frac{3 J_{16}}{2 s},\quad 
    J_{18}'=-\frac{3 y J_{16}}{2(s-1)}\,.
    \label{eq:egeeDE}    
\end{gather} 
The boundary conditions follow from the threshold asymptotics $s\to 9$. Moreover, for integrals $J_{3-18}$ it is sufficient to know that they vanish in this limit. In particular, using the last line of the above formula we have $J_{13}=J_{14}=J_{15}=J_{16}=J_{17}=J_{18}=0$. This means that the $J_{1-18}$ do not form a complete $\e$-regular basis, but the subset $J_{1-12}$ does, as far as the total cross section is concerned.

Note that, apart from the first two equations, the differential system is strictly triangular and the integrals $J_{3-12}$ can be elementary expressed as the iterated integrals depending on $J_1$ by integrating the corresponding equation. Alternatively, similar to Ref. \cite{Lee2019}, we can express them as one-fold integrals of $J_1$ and (poly)logarithms. We have 
\begin{gather}
    J_1(s)=\frac{s-9}{6 \left(\sqrt{s}-1\right)^{3/2} \left(\sqrt{s}+3\right)^{1/2}}\mathrm{K}\left(\frac{\left(\sqrt{s}-3\right) \left(\sqrt{s}+1\right)^3}{\left(\sqrt{s}-1\right)^3 \left(\sqrt{s}+3\right)}\right),\nonumber\\
    J_2(s)=\frac{1}{8s} \left(\sqrt{s}-1\right)^{3/2} \left(\sqrt{s}+3\right)^{1/2}\, \mathrm{E}\left(\frac{\left(\sqrt{s}-3\right) \left(\sqrt{s}+1\right)^3}{\left(\sqrt{s}-1\right)^3 \left(\sqrt{s}+3\right)}\right)
    -\frac{2J_1(s)}{ \sqrt{s}(\sqrt{s}-3)},\nonumber\\
    J_3(s)=2\intop_9^s d\tilde{s}\frac{J_1(\tilde{s})}{\tilde{s}-9},\quad
    J_4(s)=2\intop_9^s d\tilde{s}\frac{J_1(\tilde{s})}{\tilde{s}-9}\ln\left(\frac{s-1}{\tilde{s}-1}\right),\quad
    J_5(s)=\intop_9^s d\tilde{s}\frac{J_1(\tilde{s})}{\tilde{s}-1},\nonumber\\
    J_6(s)=\frac12\intop_9^s d\tilde{s}\frac{J_1(\tilde{s})}{\tilde{s}-1}\ln \left(\frac{s}{\tilde{s}}\right),\quad
    J_7(s)=-3\intop_9^s d\tilde{s}\frac{J_1(\tilde{s})}{\tilde{s}-9}\ln \left(\frac{s}{\tilde{s}}\right),\nonumber\\
    J_8(s)=2\intop_9^s d\tilde{s} \frac{(7\tilde{s}-15)J_1(\tilde{s})}{\left(\tilde{s}-9\right) \left(\tilde{s}-1\right)}
    \ln\left(\frac{\sqrt{s-1}+\sqrt{s+3}}{\sqrt{\tilde{s}-1}+\sqrt{\tilde{s}+3}}\right),\quad
    J_9(s)=\intop_9^s d\tilde{s} 
    \frac{\frac{7\tilde{s}-15}{\tilde{s}-9}J_1(\tilde{s})}{\sqrt{(\tilde{s}-5)(\tilde{s}-1)}},\nonumber\\
    J_{10}(s)=\intop_9^s d\tilde{s} \frac{\frac{7 \tilde{s}-15}{\tilde{s}-9}J_1(\tilde{s})}{\sqrt{(\tilde{s}-5)(\tilde{s}-1)}} \ln\left(\frac{s-1}{\tilde{s}-1}\right)
    -32\intop_9^s d\tilde{s} \frac{\tilde{s}J_1(\tilde{s})}{(\tilde{s}-9)(\tilde{s}-1)} \ln \left(\frac{\sqrt{s-5}+\sqrt{s-1}}{\sqrt{\tilde{s}-5}+\sqrt{\tilde{s}-1}}\right),\nonumber\\
    J_{11}(s)=\intop_9^s d\tilde{s} \frac{J_1(\tilde{s})}{\sqrt{(\tilde{s}-5)(\tilde{s}-1)}} \frac{7 \tilde{s}-15}{\tilde{s}-9} \ln \left(\frac{\sqrt{s-5}+\sqrt{s-1}}{\sqrt{\tilde{s}-5}+\sqrt{\tilde{s}-1}}\right),\nonumber\\
    J_{12}(s)=\intop_9^s d\tilde{s} \frac{\frac{7 \tilde{s}-15}{\tilde{s}-9}J_1(\tilde{s})}{\sqrt{(\tilde{s}-5)(\tilde{s}-1)}} \ln \left(\frac{s-5}{\tilde{s}-5}\right)
    \raggedright{+2\intop_9^s d\tilde{s} \frac{(5 \tilde{s}+3)J_1(\tilde{s})}{\left(\tilde{s}-9\right) \left(\tilde{s}-1\right)} \ln \left(\frac{\sqrt{s-5}+\sqrt{s-1}}{\sqrt{\tilde{s}-5}+\sqrt{\tilde{s}-1}}\right)\,.}\label{eq:Jes}
\end{gather}

Taking the traces over $\gamma$-matrices, performing the IBP reduction and expressing the result via $J_k$, we obtain for the total cross section
\begin{multline}\label{eq:egeee}
    \sigma_{\egeee}=\frac{\alpha ^3}{s-1} 
    \bigg[
    \frac{4\left(168 s^4-2714 s^3+8697 s^2+1128 s+81\right) J_1}{27(s-1)^2 s}\\
    -\frac{2 \left(872 s^4-11289 s^3+29949 s^2+729s+27\right) J_2}{27 (s-1)^2 s}
    -\frac{32 \left(s^2-5 s-2\right)J_6}{(s-1)^2}\\
    +\frac{4 \left(24 s^6-38s^5-228 s^4+507 s^3+1161 s^2-45 s+27\right) J_3}{3 (s-1)^2s^2 (s+3)}
    -\frac{16 \left(9 s^2-46 s+17\right)J_4}{(s-1)^2}\\
    -\frac{8 \left(3 s^6-3 s^5-59 s^4-94 s^3-159 s^2+33s-9\right) J_5}{3 (s-1)^2 s^2 (s+3)}
    +\frac{16 \left(2 s^2-8 s-1\right)J_7}{(s-1)^2}\\
    +\frac{32 \left(s^4-s^3-21 s^2-27s+16\right) y J_8}{(s-1)^3 (s+3)}
    -\frac{8 \left(3 s^4-30s^3+116 s^2-146 s-7\right) z J_9}{3 (s-5) (s-1)^2}\\
    -\frac{8 (s+1) z J_{10}}{s-1}
    +\frac{8 \left(3 s^2-6 s-5\right)J_{11}}{(s-1)^2}
    -\frac{16 \left(s^3-3 s^2-9 s+19\right) zJ_{12}}{(s-5) (s-1)^2}
    \bigg]\,.
\end{multline}
The above formula, together with Eq. \eqref{eq:Jes}, represents our exact result for the total cross section.

\subsection{Asymptotics}
Let us now derive the asymptotics of the obtained results. 
\paragraph{Threshold asymptotics} The threshold asymptotics can be easily obtained by expanding under the integral sign. Taking into account the position of singularities in the differential system \eqref{eq:egeeDE}, we choose to expand in the variable $r=\frac{s-9}{s-1}$. This choice secures the fast convergence of power series for any $s>9$.

We have
\begin{equation}
    \sigma_{\egeee}=\frac{\pi}{3 \sqrt{3}}  \left[\frac{4 r^2}{27}+\frac{602 r^3}{243}+\frac{4217 r^4}{4374}+O(r^5)\right].
\end{equation}
The leading term is in agreement with the asymptotics found in Ref. \cite{votruba1948pair}.

\begin{figure}
    \centering
    \includegraphics[width=0.6\linewidth]{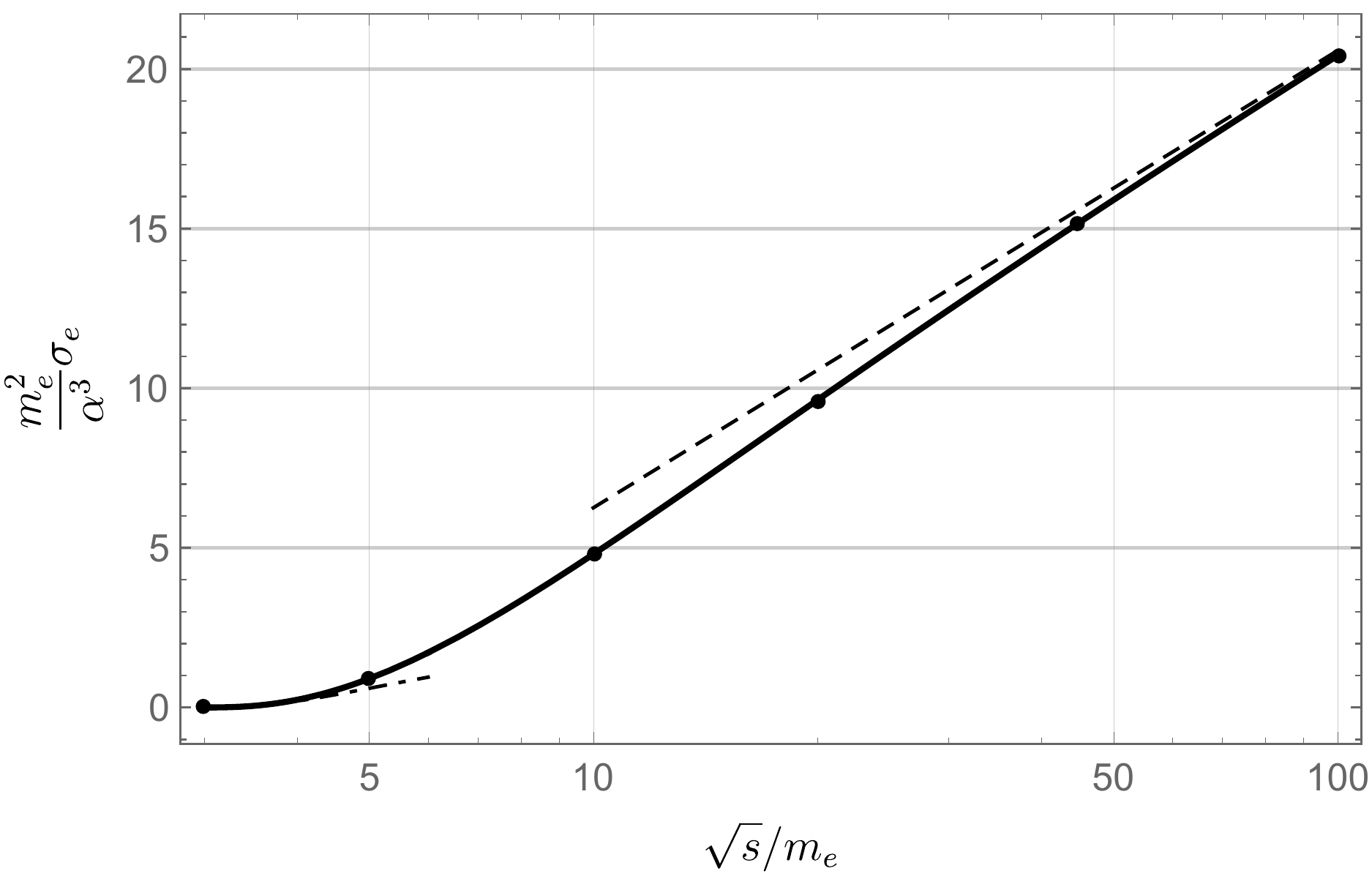}
    \caption{Exact cross section (solid curve) and asymptotics. The high-energy asymptotics (dotted line) corresponds to the leading term $\alpha ^3 \left(\frac{28 \ln{s}}{9}-\frac{218}{27}\right)$ only, while Eq. \eqref{eq:HEeee}, taking into account the next-to-leading term, would lead to the curve hardly discernible from the exact result: e.g., at $\sqrt{s}=10m_e$ the difference is a tiny $0.4$\%. Points show the numerical results of Ref. \cite{haug1975bremsstrahlung}.}
    \label{fig:cs}
\end{figure}

\paragraph{High-energy asymptotics} The high-energy asymptotics of $J_1$ and $J_2$ can be calculated directly:
\begin{align}
    J_1(s)&\stackrel{s\to\infty}{\sim}    
    \frac{\ln{s}}{8}-\frac{3\ln{s}+2}{4 s}-\frac{6 \ln{s}-5}{4 s^2}+O\left(s^{-3}\right),\\
    J_2(s)&\stackrel{s\to\infty}{\sim} \frac{1}{8}-\frac{2 \ln{s}+3}{8 s} -\frac{3 \ln{s}-1}{4 s^2}+O\left(s^{-3}\right)\,.
\end{align}
However, the calculation of the asymptotics of $J_{3-12}$ is more tricky. 
Let us explain our approach on the example of $J_3(s)$. We transform it in the following way:
\begin{multline}
    J_3(s)=2\intop_9^s d\tilde{s}\frac{J_1(\tilde{s})}{\tilde{s}-9}=2\intop_9^s d\tilde{s}\left[\frac{J_1(\tilde{s})}{\tilde{s}-9}-\frac{\ln \tilde{s}}{8\tilde{s}}\right]+\frac{\ln^2s-\ln^29}{8}\\
    =\frac{\ln^2s-\ln^29}{8}+C_3-2\intop_s^\infty d\tilde{s}\left[\frac{J_1(\tilde{s})}{\tilde{s}-9}-\frac{\ln \tilde{s}}{8\tilde{s}}\right]\,,\label{eq:ee:Jasym}
\end{multline}
where $C_3=2\int_9^\infty d\tilde{s}\left[\frac{J_1(\tilde{s})}{\tilde{s}-9}-\frac{\ln \tilde{s}}{8\tilde{s}}\right]$. The first transition in Eq. \eqref{eq:ee:Jasym} is the subtraction from the integrand of its large-$\tilde{s}$ asymptotics so as to secure the convergence of the integral when its upper limit goes to $\infty$. Now, in the last term we can substitute the asymptotics of the integrand and integrate term-wise. The only nontrivial task left is to evaluate $C_3$. While we were not able to find an analytic way to perform this integration, we have successfully applied \texttt{PSLQ} algorithm \cite{ferguson1991polynomial} to find that $C_3=\frac18\ln^2 9-\frac{\pi ^2}{24}$. Thus, we obtain
\begin{equation}
    J_3(s)=\frac{\ln ^2{s}}{8}-\frac{\pi ^2}{24}+\frac{1-3 \ln {s}}{4s}+\frac{37-30 \ln{s}}{16s^2}+\ldots
\end{equation}
Note that we could have obtained as many terms of the asymptotics as needed. Similarly, we have found the asymptotics of all integrals.
Finally we obtain
\begin{multline}\label{eq:HEeee}
    \sigma_{\egeee}=\alpha ^3 \bigg[\frac{28 \ln{s}}{9}-\frac{218}{27}\\
    -\frac{1}{s}\left(\frac{8}{3} \ln^3{s}-\frac{49 }{6}\ln ^2{s}-\left(2 \pi ^2-\frac{409}{9}\right)\ln{s}+12 \zeta_3+\frac{31 \pi^2}{18}-\frac{1081}{12}\right)+O\left(s^{-2}\right)\bigg]
\end{multline}
The leading term $\alpha ^3 \left(\frac{28 \ln{s}}{9}-\frac{218}{27}\right)$ agrees with Ref. \cite{Baier}. The subleading term has been considered in Ref. \cite{haug1981simple}. In that paper the expansion has been performed with respect to photon energy $k=(s-1)/2$ in the rest frame of the initial electron. If  we express Eq. \eqref{eq:HEeee} in terms of $k$, we obtain
\begin{multline}\label{eq:HEeee1}
    \sigma_{\egeee}=\alpha ^3 \bigg[\frac{28}{9} \ln (2 k)-\frac{218}{27}\\
    +\frac{1}{k}\left(-\frac{4}{3}  \ln ^3(2 k)+\frac{49}{12} \ln ^2(2 k)+\left(\pi ^2 -\frac{409}{18}\right) \ln (2 k)-6 \zeta_3-\frac{31 \pi ^2}{36}+\frac{3355}{72}\right)+O\left(k^{-2}\right)\bigg]\\
    \approx
    \alpha ^3 \bigg[\frac{28}{9} \ln (2 k)-\frac{218}{27}-\frac{1}{k}\left(\frac{4}{3}  \ln ^3(2 k)-4.08\ln ^2(2 k)+12.85 \ln (2 k)-30.89\right)+O\left(k^{-2}\right)\bigg].
\end{multline}
This is to be compared with the result of Ref. \cite{haug1981simple}:
\begin{multline}\label{eq:HEeee2}
    \sigma^{\text{Ref. \cite{haug1981simple}}}_{\egeee}
    \approx
    \alpha ^3 \bigg[\frac{28}{9} \ln (2 k)-\frac{218}{27}-\frac{1}{k}\left(\frac{4}{3}  \ln ^3(2 k)-3.86\ln ^2(2 k)+11 \ln (2 k)-27.9\right)\bigg].
\end{multline}
Figure  \ref{fig:cs} shows the cross section as a function of energy. It is remarkable that the two first terms of high-energy asymptotics, Eq. \eqref{eq:HEeee}, provide a very good approximation of the total cross section with accuracy better than $1$\% for $\sqrt{s}>8m_e$. In the same figure we have presented a few points of Ref. \cite{haug1975bremsstrahlung} obtained by numerical integration. These points serve as a perfect cross check of our analytic results.

\section{Conclusion}

In the present paper we have calculated the total cross sections for the processes $\egeXX$ with $X=\mu^{-},\gamma,e^-$ for arbitrary energies. Apart from the last case, our exact results are expressed via classical polylogarithms. We have calculated the threshold and high-energy asymptotics of the obtained results and compared them with the available results finding some discrepancies in the latter.

%

\begin{figure}[ht]
    \begin{subfigure}[t]{.32\textwidth}
        \centering
        \includegraphics[width=1\linewidth]{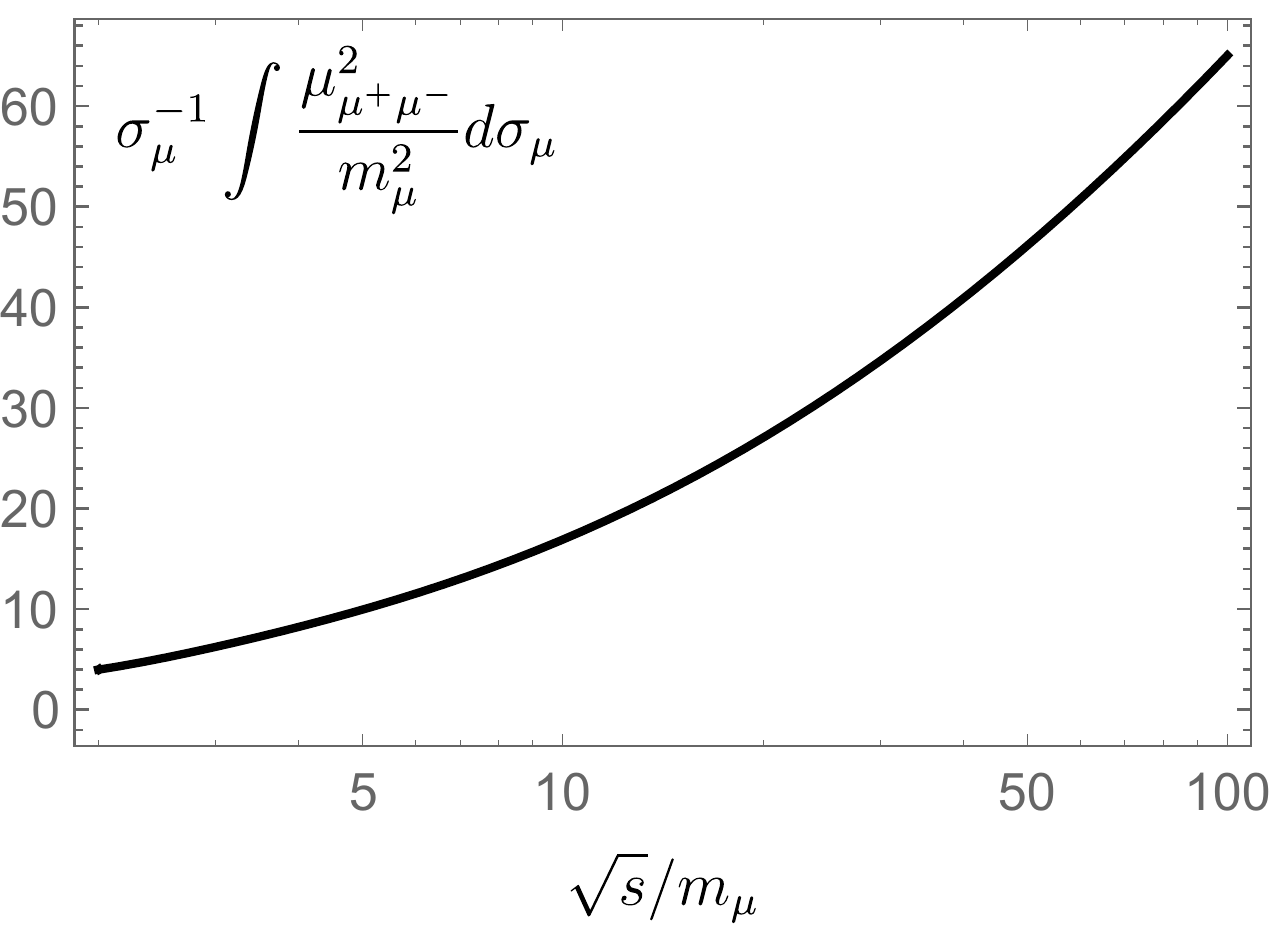}
        \caption{Cross section averaged square of the $\mu^+\mu^-$ pair invariant mass, in units $m_\mu^2$, as function of $\sqrt{s}/m_{\mu}$.}\label{fig:muweighted}
    \end{subfigure}
    \hfill
    \begin{subfigure}[t]{.325\textwidth}
        \centering
        \includegraphics[width=1\linewidth]{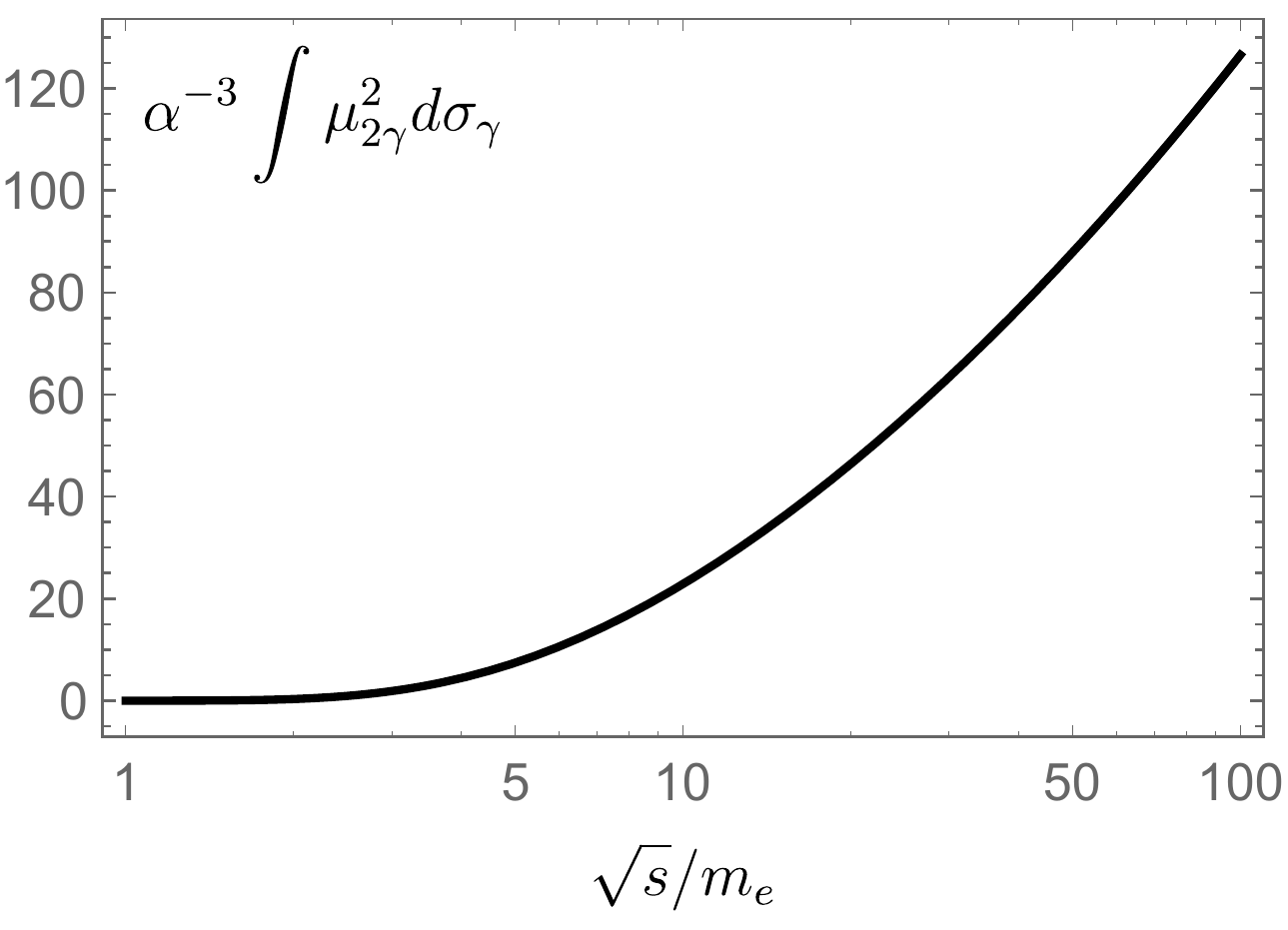}
        \caption{Double Compton scattering cross section weighted by the square of invariant mass of photons as function of $\sqrt{s}/m_{\mu}$.}\label{fig:gammaweighted}
    \end{subfigure}
    \hfill
    \begin{subfigure}[t]{.32\textwidth}
        \centering
        \includegraphics[width=1\linewidth]{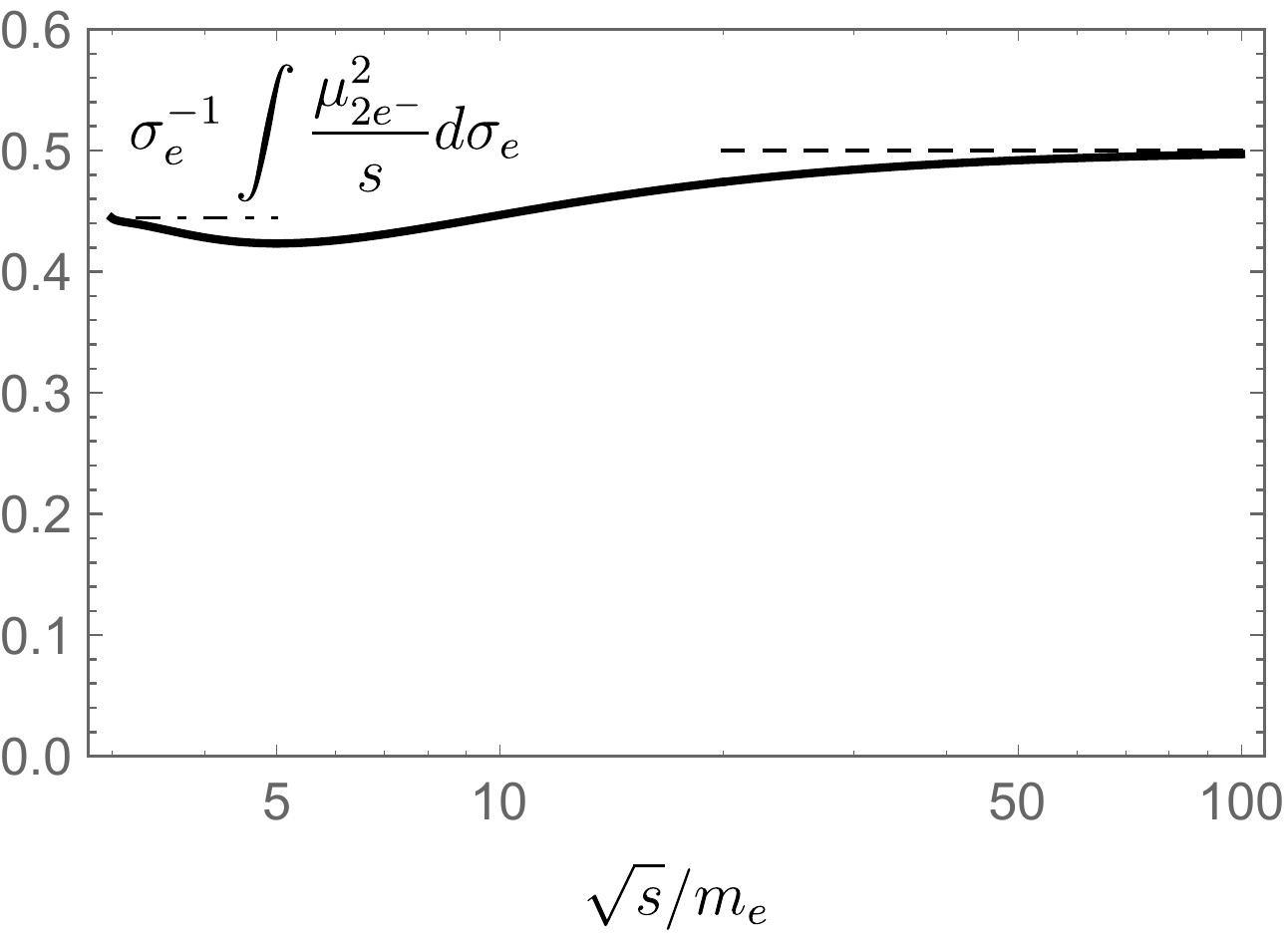}
        \caption{Cross section averaged square of $e^-e^-$ invariant mass, divided by $s$, as function of $\sqrt{s}/m_{\mu}$. The low- and high-energy asymptotics are $4/9$ and $1/2$, respectively.}\label{fig:eweighted}
    \end{subfigure}
    \caption{Weighted cross sections defined in Eq. \eqref{eq:weighted}.}
    \label{fig:weighted}
\end{figure} 
Note that our method allows one to obtain exact expressions also for the integrated cross sections weighted by some kinematic invariants. As an example, we have calculated the cross sections weighted by the invariant mass of a pair of produced particles for each process. More precisely, we consider
\begin{align}\label{eq:weighted}
    \langle\mu_{\mu^+\mu^-}^2\rangle&\stackrel{\text{def}}{=}\int \mu_{\mu^+\mu^-}^2 d\sigma_{\egemumu}, \nonumber\\ 
    \langle\mu_{2\gamma}^2\rangle&\stackrel{\text{def}}{=}\int \mu_{2\gamma}^2 d\sigma_{\egegg},\nonumber\\
    \langle\mu_{2e^-}^2\rangle&\stackrel{\text{def}}{=}\int \mu_{2e^-}^2  d\sigma_{\egeee}.
\end{align}
i.e., we consider the cross section weighted by the square of invariant mass of the muon pair, the photon pair, and the electron pair  for \egemumu, \egegg, and \egeee, respectively. Note that $\int \mu_{\gamma\gamma}^2  d\sigma_{\egegg}$ is infrared safe. The numerical results are presented in Fig. \ref{fig:weighted}. We provide the explicit expressions for the weighted  cross sections in Eq. \eqref{eq:weighted} in ancillary file, see below. 
Let us note that the leading asymptotics of $\langle\mu_{2e^-}^2\rangle$ is equal to that of $\sigma_{\egeee}$ multiplied by $s/2$. This can be easily understood on physical ground as follows. At high energies the electron-positron pair is mostly produced with small invariant mass by equivalent photon mechanism. Then we have $p_3\approx p_4$, $p_2\approx (\sqrt{s}/2,\boldsymbol{n}_2\sqrt{s}/2)$, and  $p_3+p_4\approx 2p_3\approx q(\sqrt{s}/2,-\boldsymbol{n}_2\sqrt{s}/2)$. Then $\mu_{2e^-}^2=(p_2+p_3)^2\approx2p_2\cdot p_3\approx s/2$.

For reader convenience, we attach to the e-print submission the ancillary file \texttt{CSall.m} which contains evaluation-ready expressions for the exact cross sections defined in Eqs. \eqref{eq:cs egemumu}, \eqref{eq:cs egepipi}, \eqref{eq:egeggcsrf}, \eqref{eq:egeggcscmf}, \eqref{eq:egeee}, and \eqref{eq:weighted}. When read in from \textit{Mathematica} session with \texttt{Get["CSall.m"]} the file prints all necessary information about the functions defined therein.

The last but not the least, the calculation presented in this paper has demonstrated a few methods not so widely known in the multiloop community. First, in the calculation of the \egemumu cross section we have shown how to obtain the differential equations with respect to one variable ($s$ in our case) for the coefficients of generalized power series with respect to another variable ($m_e^2$ in our case). Next, we have explicitly demonstrated that the contribution of soft photons to the integrated cross section can be calculated by means of the multiloop techniques. Finally, we have successfully applied the  recently introduced \cite{Lee2019} approach to the non-polylogarithmic integrals, based on the construction of $\e$-regular basis. These approaches can be applied in other physically relevant calculations.

\paragraph*{Acknowledgments} This work has been supported by Russian Science Foundation, grant 20-12-00205.


\begin{thebibliography}{10}
    
    \bibitem{Racah1934a}
    G.~Racah, \emph{Sulla nascita degli elettroni positivi},
    \href{https://doi.org/10.1007/BF02959919}{\emph{Il Nuovo Cimento (1924-1942)}
        {\bfseries 11} (1934) 477}.
    
    \bibitem{Racah1934}
    G.~Racah, \emph{Sopra l’irradiazione nell’urto di particelle veloci},
    \href{https://doi.org/10.1007/BF02959918}{\emph{Il Nuovo Cimento} {\bfseries
            11} (1934) 461}.
    
    \bibitem{Remiddi:1999ew}
    E.~Remiddi and J.~A.~M. Vermaseren, \emph{{Harmonic polylogarithms}},
    \href{https://doi.org/10.1142/S0217751X00000367}{\emph{Int. J. Mod. Phys.}
        {\bfseries A15} (2000) 725}
    [\href{https://arxiv.org/abs/hep-ph/9905237}{{\ttfamily hep-ph/9905237}}].
    
    \bibitem{Goncharov1998}
    A.~B. Goncharov, \emph{Multiple polylogarithms, cyclotomy and modular
        complexes}, {\emph{Mathematical Research Letters} {\bfseries 5} (1998) 497}.
    
    \bibitem{1991A&A...252..414D}
    C.~D. {Dermer} and R.~{Schlickeiser}, \emph{{Effects of triplet pair production
            on ultrarelativistic electrons in a soft photon field.}}, {\emph{Astronomy \&
            Astrophysics} {\bfseries 252} (1991) 414}.
    
    \bibitem{ENDO1993517}
    I.~Endo and T.~Kobayashi, \emph{Exact evaluation of triplet photoproduction},
    \href{https://doi.org/https://doi.org/10.1016/0168-9002(93)90669-9}{\emph{Nuclear
            Instruments and Methods in Physics Research Section A: Accelerators,
            Spectrometers, Detectors and Associated Equipment} {\bfseries 328} (1993) 517
    }.
    
    \bibitem{10.1093/mnras/266.4.910}
    A.~Mastichiadis, R.~J. Protheroe and A.~P. Szabo, \emph{{The effect of triplet
            production on pair–Compton cascades in thermal radiation}},
    \href{https://doi.org/10.1093/mnras/266.4.910}{\emph{Monthly Notices of the
            Royal Astronomical Society} {\bfseries 266} (1994) 910}
    [\href{https://arxiv.org/abs/https://academic.oup.com/mnras/article-pdf/266/4/910/3051921/mnras266-0910.pdf}{{\ttfamily
            https://academic.oup.com/mnras/article-pdf/266/4/910/3051921/mnras266-0910.pdf}}].
    
    \bibitem{PhysRevLett.86.1430}
    A.~Kusenko and M.~Postma, \emph{Neutrinos produced by ultrahigh-energy photons
        at high redshift},
    \href{https://doi.org/10.1103/PhysRevLett.86.1430}{\emph{Phys. Rev. Lett.}
        {\bfseries 86} (2001) 1430}.
    
    \bibitem{PhysRevD.64.071302}
    H.~Athar, G.-L. Lin and J.-J. Tseng, \emph{Muon pair production by
        electron-photon scatterings},
    \href{https://doi.org/10.1103/PhysRevD.64.071302}{\emph{Phys. Rev. D}
        {\bfseries 64} (2001) 071302}.
    
    \bibitem{haug2004pair}
    E.~Haug, \emph{Pair production by photons in a hot maxwellian plasma},
    {\emph{Astronomy \& Astrophysics} {\bfseries 416} (2004) 437}.
    
    \bibitem{Ravenni:2020ven}
    A.~Ravenni and J.~Chluba, \emph{{The double Compton process in astrophysical
            plasmas}},  \href{https://arxiv.org/abs/2005.06941}{{\ttfamily 2005.06941}}.
    
    \bibitem{votruba1948pair}
    V.~Votruba, \emph{Pair production by $\gamma$-rays in the field of an
        electron}, {\emph{Physical Review} {\bfseries 73} (1948) 1468}.
    
    \bibitem{mork1967pair}
    K.~J. Mork, \emph{Pair production by photons on electrons}, {\emph{Physical
            Review} {\bfseries 160} (1967) 1065}.
    
    \bibitem{PhysRevA.4.917}
    K.~J. Mork, \emph{Radiative corrections. ii. compton effect},
    \href{https://doi.org/10.1103/PhysRevA.4.917}{\emph{Phys. Rev. A} {\bfseries
            4} (1971) 917}.
    
    \bibitem{Baier}
    V.~N. Baier, V.~M. Katkov and V.~S. Fadin, \emph{Radiation of relativistic
        electrons}. 1, 1973.
    
    \bibitem{haug1981simple}
    E.~Haug, \emph{Simple analytic expressions for the total cross section for
        $\gamma$-e pair production}, {\emph{Zeitschrift f{\"u}r Naturforschung A}
        {\bfseries 36} (1981) 413}.
    
    \bibitem{gould1984cross}
    R.~Gould, \emph{The cross section for double compton scattering}, {\emph{The
            Astrophysical Journal} {\bfseries 285} (1984) 275}.
    
    \bibitem{anguelov1999numerical}
    V.~Anguelov, S.~Petrov, L.~Gurdev and J.~Kourtev, \emph{On the numerical
        analysis of triplet pair production cross-sections and the mean energy of
        produced particles for modelling electron-photon cascade in a soft photon
        field}, {\emph{Journal of Physics G: Nuclear and Particle Physics} {\bfseries
            25} (1999) 1733}.
    
    \bibitem{Lee2019}
    R.~N. Lee and A.~I. Onishchenko, \emph{$\epsilon$-regular basis for
        non-polylogarithmic multiloop integrals and total cross section of the
        process $e^+e^-\to 2(q\bar q)$},
    \href{https://arxiv.org/abs/1909.07710v2}{{\ttfamily 1909.07710v2}}.
    
    \bibitem{Lee2013a}
    R.~N. Lee, \emph{{LiteRed 1.4: a powerful tool for reduction of multiloop
            integrals}}, \href{https://doi.org/10.1088/1742-6596/523/1/012059}{\emph{J.
            Phys. Conf. Ser.} {\bfseries 523} (2014) 012059}
    [\href{https://arxiv.org/abs/1310.1145}{{\ttfamily 1310.1145}}].
    
    \bibitem{Lee2018}
    R.~N. Lee, A.~V. Smirnov and V.~A. Smirnov, \emph{Solving differential
        equations for feynman integrals by expansions near singular points},
    \href{https://doi.org/10.1007/JHEP03(2018)008}{\emph{JHEP} (2018) }.
    
    \bibitem{Henn2013}
    J.~M. Henn, \emph{{Multiloop integrals in dimensional regularization made
            simple}},
    \href{https://doi.org/10.1103/PhysRevLett.110.251601}{\emph{Phys.Rev.Lett.}
        {\bfseries 110} (2013) 251601}
    [\href{https://arxiv.org/abs/1304.1806}{{\ttfamily 1304.1806}}].
    
    \bibitem{Libra}
    R.~N. Lee, ``\texttt{Libra}, a tool for reducing differential systems to
    $\epsilon$-form.''.
    
    \bibitem{hahn2016concurrent}
    T.~Hahn, \emph{Concurrent {C}uba}, {\emph{Computer Physics Communications}
        {\bfseries 207} (2016) 341}.
    
    \bibitem{ram1971calculation}
    M.~Ram and P.~Wang, \emph{Calculation of the total cross section for double
        compton scattering}, {\emph{Physical Review Letters} {\bfseries 26} (1971)
        476}.
    
    \bibitem{lotstedt2013theoretical}
    E.~L{\"o}tstedt and U.~D. Jentschura, \emph{Theoretical study of the compton
        effect with correlated three-photon emission: From the differential cross
        section to high-energy triple-photon entanglement}, {\emph{Physical Review A}
        {\bfseries 87} (2013) 033401}.
    
    \bibitem{ferguson1991polynomial}
    H.~R. Ferguson and D.~H. Bailey, \emph{A polynomial time, numerically stable
        integer relation algorithm}, .
    
    \bibitem{haug1975bremsstrahlung}
    E.~Haug, \emph{Bremsstrahlung and pair production in the field of free
        electrons}, {\emph{Zeitschrift f{\"u}r Naturforschung A} {\bfseries 30}
        (1975) 1099}.
    
\end{thebibliography}

\providecommand{\href}[2]{#2}\begingroup\raggedright\endgroup

\end{document}